\documentclass[twocolumn,prb,aps]{revtex4-2}
\usepackage{amsthm,amsmath,amssymb,lipsum,mathrsfs}
\usepackage{graphicx}
\usepackage[T1]{fontenc}

\begin{document}

\title{Fourfold Anisotropic Magnetoresistance and Unconventional Critical Exponents in Twinned FePd$_2$Te$_2$}

\author{Zhaoxu Chen}%
\affiliation{Beijing National Laboratory  for Condensed Matter Physics, Institute of Physics, Chinese Academy of Sciences, Beijing 100190, China}
\affiliation{School of Physical Sciences, University of Chinese Academy of Sciences, Beijing 100049, China}
\author{Yuxin Yang}%
\affiliation{Beijing National Laboratory  for Condensed Matter Physics, Institute of Physics, Chinese Academy of Sciences, Beijing 100190, China}
\affiliation{College of Materials Sciences and Opto-Electronic Technology, University of Chinese Academy of Sciences, Beijing 100049, China}
\author{Jian-gang Guo}%
\email{jgguo@iphy.ac.cn}
\affiliation{Beijing National Laboratory  for Condensed Matter Physics, Institute of Physics, Chinese Academy of Sciences, Beijing 100190, China}

\begin{abstract}
As a special material symmetry operation, crystal twins usually influence physical properties. Here, detailed electrical transport and magnetic measurements were performed to reveal twinning effect on properties of van der Waals ferromagnet FePd$_2$Te$_2$. Orthorhombic crystal domains were observed in polarized optical microscopy and fixed $\pi/2$ angle between adjacent domains suggests a phase transition origin of the twins. FePd$_2$Te$_2$ exhibits fourfold in-plane anisotropic magnetoresistance. It is attributed to antiferromagnetic coupling component near atomically flat twin boundary and pseudo four-fold symmetry from perpendicular Fe chains. Intense magnetic domain motion is suggested by Hopkinson effect observed in magnetic susceptibility. A set of unusual critical exponents $\beta = 0.866$, $\gamma = 1.043$, $\delta = 2.20$ cannot be classified in any universal class predicted by renormalized group. Deviation from standard model reflects the non-saturating magnetization and slow growth of spontaneous magnetization resulting from crystal domain walls. These results show that additional symmetry from twins and twin boundary have a significant effect on electrical transport and magnetic properties of FePd$_2$Te$_2$. There is much room to modulate physical properties of twinned van der Waals ferromagnets through twins.
\end{abstract}
\maketitle
\section{Introduction}
A twinned crystal is an aggregate in which domains with different orientations growth together according to symmetry operation described by twin laws\cite{parsons2003introduction}. Crystal twins can be generated by a solid state phase transition, an external force or a perturbation during the crystal growth\cite{nespolo2015tips}. For crystal twins generated by the latter two random factors, the symmetry relation between different domains cannot be given a prior orientation\cite{nespolo2015tips}. Whereas, phase-transition-driven crystal twinning effect is more interesting and has been studied deeply in strong correlated systems\cite{fisher2011plane}. For example, in ferroelastic-like tetragonal-to-orthorhombic transition during cooling, $C_4$ is lost as a point group symmetry operation. To minimize the elastic and release strain, twin domain structures form. Such orthorhombic domains have been widely observed in unconventional superconductors and related materials with tetragonal-to-orthorhombic phase transformation, such as YBa$_2$Cu$_3$O$_{7-\delta}$\cite{specht1988effect}, La$_{2-x}$Sr$_x$CuO$_4$\cite{radaelli1994structural}, AFe$_2$As$_2$ (A = Ca/Sr/Ba/Eu)\cite{tanatar2009direct} and Co-doped counterpart\cite{prozorov2009intrinsic}, NaFeAs\cite{li2009structural} and FeSe\cite{mcqueen2009tetragonal}.

Although crystal twins is only a structural concept, it has significant effect on physical properties. In superconductors, as a kind of defects, domain walls between adjacent twins can effectively pin the vortices and increase the critical current density\cite{dolan1989vortex,maggio1997critical,song2012suppression}. In terms of superconducting mechanisms, twin boundaries can invoke twist of the order parameter structure and breaks time-reversal symmetry\cite{tsuei2000pairing,watashige2015evidence}. Besides, in Fe-based superconductors, coupling between twins and magnetism has been deeply studied in the anisotropy of in-plane resistivity, charge dynamics and electronic structure\cite{fisher2011plane}. Compared with these strong correlated materials and conventional magnetic films, due to the lack of twinned bulk counterpart, study on interplay between twinning effect and van der Waals ferromagnet is not carried out yet. Recent discovered van der Waals ferromagnet FePd$_2$Te$_2$ with $T_c$ $\sim$ 180 K crystallize in a monoclinic space group $P$2$_1$/$m$ and exhibit a layered structure along [10-1] direction\cite{shi2024fepd2te2}. Within cleavage plane, perpendicular Fe chains lead to orthorhombic twins with size of $\sim$ 100 nm\cite{shi2024fepd2te2}.

Here, we report the twinning effect on electrical transport and magnetic properties of FePd$_2$Te$_2$. AC and DC magnetic susceptibility measurements reveal the paramagnetic-to-ferromagnetic transitions, which is also evidenced by anomalous Hall effect and temperature-dependent resistance. Anisotropic magnetoresistance was performed to investigate the twin-induced in-plane anisotropy and four-fold symmetry was discovered. It is noted that the in-plane anisotropic magnetoresistance can be modulated by strain introduced by extra Pd atoms. Critical exponents were calculated through detailed magnetic properties as $\beta = 0.866$, $\gamma = 1.043$, $\delta = 2.20$. These results confirmed the decouple between the dimensionality of quasi-one-dimensional Fe chain and that of mean-field-like magnetism, emphasizing the importance of crystal twins and twin boundary.
\section{Experimental methods}
FePd$_2$Te$_2$ single crystals were grown by congruent melting. Fe, Pd and Te powder were weighted with a molar ratio of 1:2:2. The starting materials were loaded into alumina crucibles and sealed in an evacuated quartz tube. The sealed quartz tube was held at 800 $^\circ C$ for 24 $h$ was cooled down to 550 $^\circ C$ within 8640 minutes. Finally, the crystals were annealed for long time and quenched in water. For FePd$_{2.08}$Te$_2$ single crystals, the growth method was similar.

X-ray diffraction was performed on a Rigaku Diffractometer with a Cu $K_\alpha$ anode (1.5418 \AA) at room temperature. The DC magnetic susceptibility and initial isothermal magnetization were measured on a Magnetic Property Measurement System (MPMS, Quantum Design). AC magnetic susceptibility was measured on Physike system. The electrical transport was measured by a Physical Property Measurement System (PPMS Dynacool, Quantum Design) and Physike TKMS based on Keithley 6221/2182A and Stanford Research System Model SR830 DSP lock-in amplifier. The angular dependence of the magnetoresistance is determined within a 270$^{\circ}$ range under a magnetic field of 9 T.

\section{Results and Discussions}
\begin{figure*}
    \centering
    \includegraphics[width=1.0\linewidth]{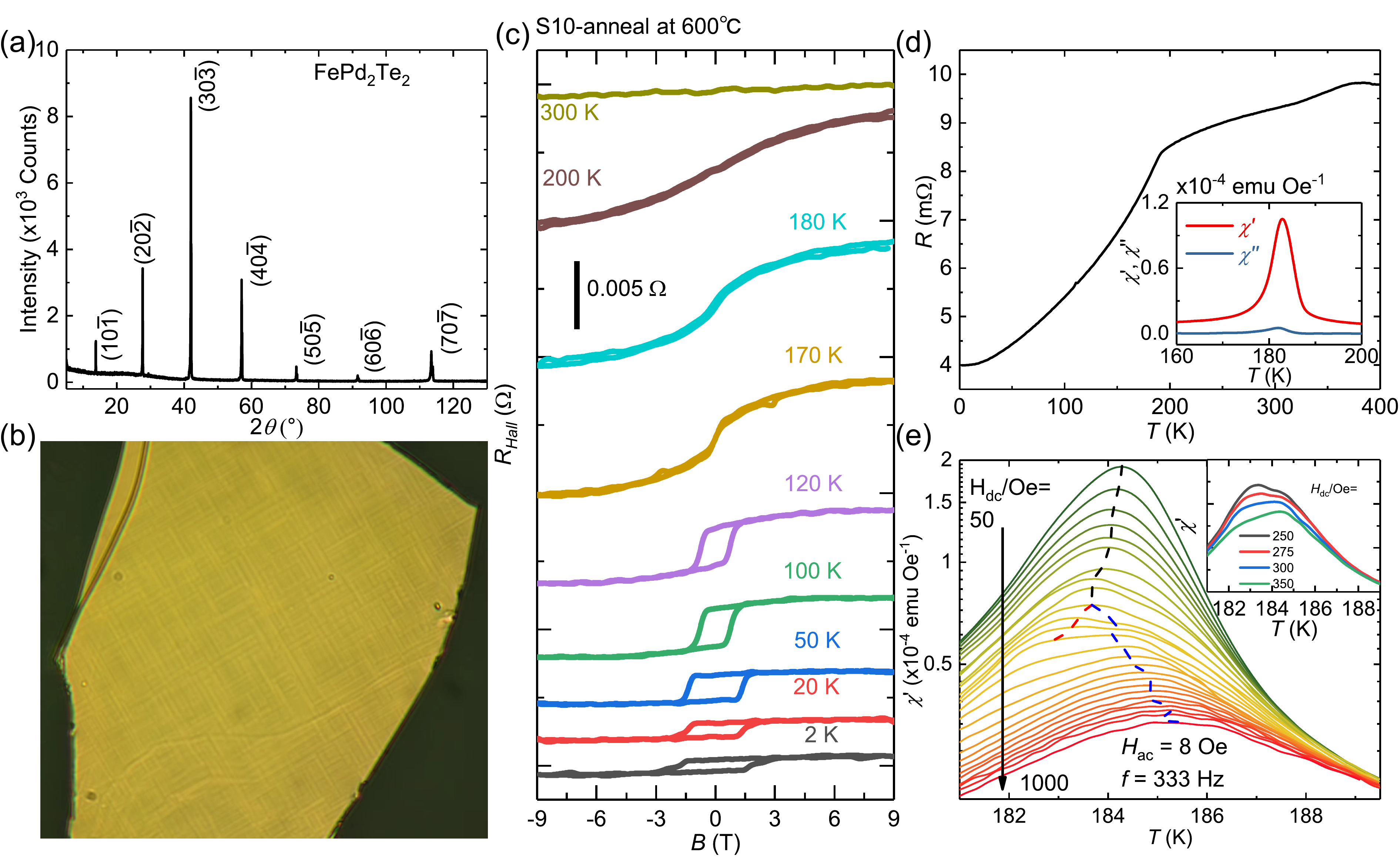}
    \caption{Structure characterization and basic physical properties. (a) X-ray diffraction of a FePd$_2$Te$_2$ single crystal. (b) Observation of orthorhombic crystal domain caused by twinning effect through optical microscopy. (c) Anomalous Hall effect of a annealed FePd$_2$Te$_2$ single crystal. (d) Temperature-dependent resistance of FePd$_2$Te$_2$ single crystal. Inset is AC magnetic susceptibility with $H_{ac}$ = 4 Oe and $f$ = 333 Hz. (e) AC magnetic susceptibility with $H_{ac}$ = 4 Oe and $f$ = 333 Hz under a serious of DC bias magnetic field from 50 Oe to 1000 Oe. Inset is enlarged plot of AC magnetic susceptibility under several selected DC bias magnetic field.}
    \label{Fig. 1}
\end{figure*}
Figure 1(a) shows the X-ray diffraction pattern of a typical annealed FePd$_2$Te$_2$ single crystal indexed by ($h$0$-h$) Miller indices. Polarized microscopy is a useful tool to observe crystal domains directly. Figure 1(b) exhibits the optical image obtained through polarized microscopy. It is noted that there are perpendicular stripes on the surface, which originate from crystal domain observed in previous scanning tunneling microscopy\cite{shi2024fepd2te2}. The twins in FePd$_2$Te$_2$ should be classified as transformation twins rather than mechanical/growth twins. Mechanical/growth twins result from random factors including external force and perturbation induced coalescence of micro-crystals, so their relative orientations are also random and cannot be predicted. Transformation twins are driven by solid state phase transition during which some specific point symmetry operations in high-symmetry phase at high temperature are lost in low-temperature phase with low symmetry, so the orientation can be predicted and are fixed\cite{nespolo2015tips}. The perpendicular crystal domains in FePd$_2$Te$_2$ indicate that there is a possible orthorhombic-to-monoclinic phase transition, which was ever observed in Bi$_2$Sr$_2$CaCu$_2$O$_y$ and need further study\cite{yang1996growth}.

As shown in Fig. 1(c), temperature-dependent of Hall effect exhibits hysteresis at low temperature, which evidences the presence of ferromagnetic phase. The paramagnetism to ferromagnetism transition is also observed in temperature-dependent resistance in Fig. 1(d). The kink indicates that the critical temperature is $\sim$ 180 K. Inset is AC magnetic susceptibility with no DC bias magnetic field. The peak corresponds to the phase transition at 184 K. It is noted that real part of AC susceptibility for a ferromagnet below $T_c$ is expected to be a constant down to low temperatures theoretically with freezing out of domain wall dynamics\cite{murthy20174f,tang1994study,benka2022interplay,karma2023study}. However, peak is also observed in many amorphous alloys\cite{drobac1996critical} and inhomogeneous phase-separated magnetic systems with small size of ferromagnetic and antiferromagnetic clusters\cite{pramanik2008phase,li2012glassy,murthy2016antisite}. It is attributed to Hopkinson effect\cite{slama2017hopkinson}, which reflects the the competition between thermo-magnetic randomization (magnetic anisotropy) and domain boundary expansion\cite{bhasker2018modulation}, suggesting the intense magnetic domain motion.

In Fig. 1(e), AC magnetic susceptibility with a serious of DC bias magnetic field from 50 to 1000 Oe was employed to investigate the magnetic properties of FePd$_2$Te$_2$. Upon increasing DC bias magnetic field, the real part peak of AC magnetic susceptibility becomes broader with reduced peak value. With zero and small DC field, the peak moves to low temperature. When the DC field reaches as high as $\sim$ 250 Oe, the peak splits into two maxima, as shown in the inset of Fig. 2(e). Further increasing DC bias field moves the first maximum to low temperature, while the other to high temperature, as indicated by the dash line. When DC bias field exceeds 450 Oe, the first peak almost vanishes. This phenomena was ever observed in other ferromagnetic systems\cite{drobac1996critical,gaunt1981critical}. The first peak is closely related to the magnetization process including domains, hysteresis, nonuniform magnetostatic modes, etc, while the second peak corresponds to the critical fluctuation\cite{gaunt1981critical}. The evolution of the first peak is in accord with that expected for Hopkinson effect, i.e., increases rapidly below $T_c$, and decreases and shifts to lower temperature with increasing small dc field\cite{drobac1996critical}. As a result, the actual peak at zero-field contains two parts of contribution from Hopkinson effect and critical fluctuation, respectively. In principle, AC susceptibility can be used to calculate critical exponents\cite{drobac1996critical}. Because it is hard to distinct the peak from critical fluctuation from Hopkinson effect, effort to analyze critical behavior from AC susceptibility was failed. 

\begin{figure*}
    \centering
    \includegraphics[width=1.0\linewidth]{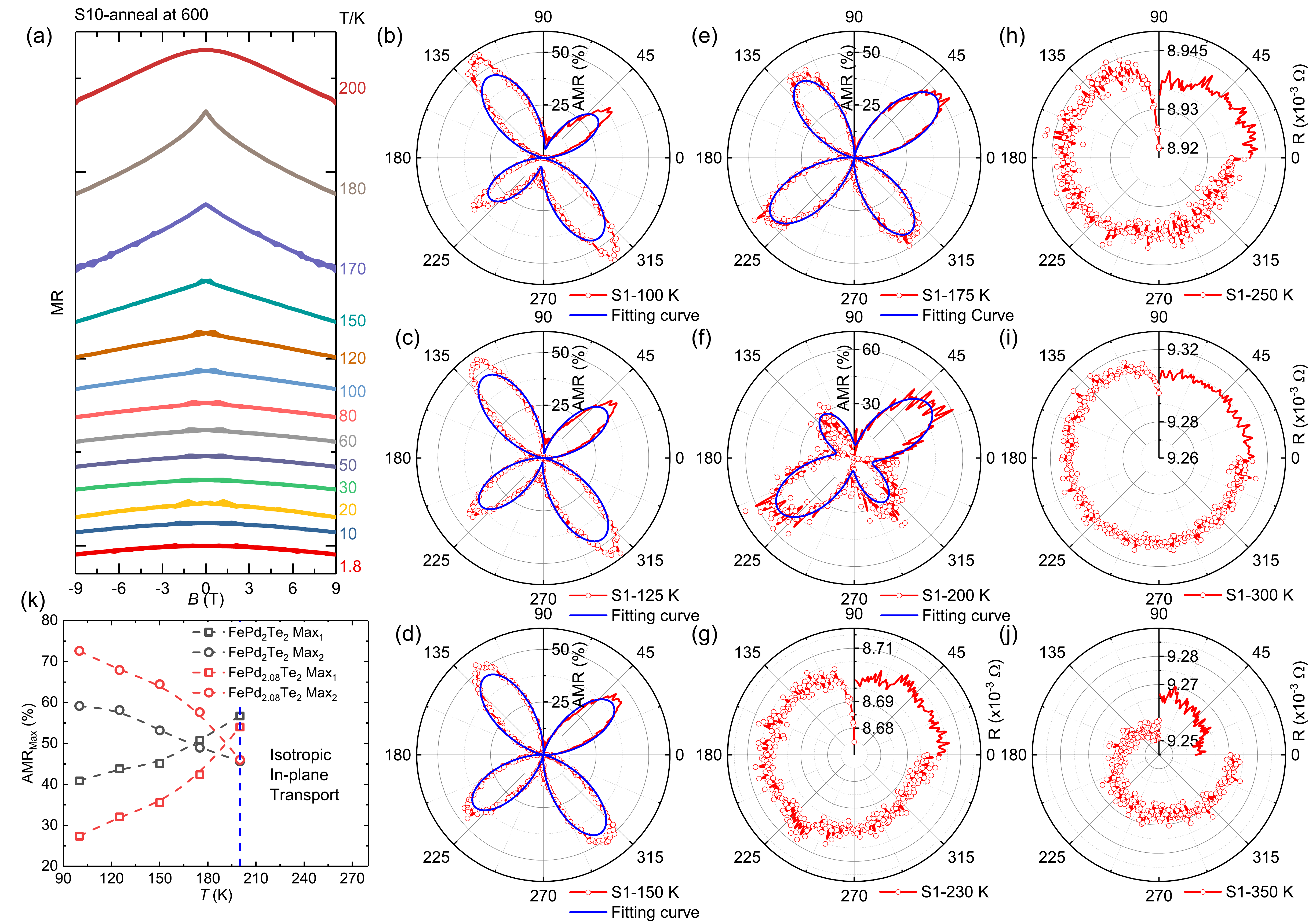}
    \caption{Anisotropic in-plane electrical transport. (a) Magnetoresistance of FePd$_2$Te$_2$ single crystal with $H$ // [10-1] from 1.8 to 200 K. (b-j) Polar plot of anisotropic in-plane magnetoresistance under a magnetic field of 9 T from 100 K to 350 K. An extra four-order term was used to fit the data. The fitting curves are blue.(k) Temperature-dependent of ratio between the two maxima in the in-plane magnetoresistance.}
    \label{Fig. 2}
\end{figure*}
Figure 2(a) shows the magnetoresistance of FePd$_2$Te$_2$ under $H$ along [10-1], i.e., out-of plane. Upon temperature decreases below $T_c$, the shape of low-field magnetoresistance change from dome-like, peak-like to butterfly-shaped hysteresis, indicating the emergence of ferromagnetism. To explore the effect of crystal twins on in-plane electrical transport, we measured in-plane anisotropic magnetoresistance under 9 T at different temperatures as shown in Fig. 2(b)-(j). There is no anisotropy within resolution at high temperature exceeding 230 K. When temperature decreases to 200 K slightly higher than $T_c$, there is a pair of strong peaks when magnetic field is perpendicular to current. Also, two weak peaks appear when magnetic field is parallel to current. Upon temperature decreases, intensity of the weak peaks become stronger and gradually exceed that of the other pair. 

Generally, anisotropic magnetoresistance is usually two-fold symmetric, which is explained by spin-orbit coupling traditionally. High resistance state emerge with $H \parallel I$, while low resistance state with $H \perp I$. Besides contribution from ordinary magnetoresistance, anisotropic magnetoresistance usually reflects the anisotropic Fermi surface and charge-spin interactions\cite{pippard1989magnetoresistance}. In magnetic materials, it can be described by
\begin{equation}
    \rho_{xx}=\rho_{\perp}+(\rho_{\parallel}-\rho_{\perp})cos^2\theta_M,
\end{equation}
 where $\theta_M$ is the angle between magnetization and current. Because obvious anisotropy starts from $\sim$ 200 K, anisotropic magnetoresistance in FePd$_2$Te$_2$ is closely related to long-range magnetic order below $T_c$ and strong magnetic fluctuation above $T_c$, respectively. It is also noted that anisotropic magnetoresistance in FePd$_2$Te$_2$ exhibit a four-fold symmetry. Such high order anisotropy magnetoresistance is also observed in other magnetic materials including magnetic films\cite{ramos2008anomalous}, antiferromagnetic cuprate with twins\cite{ando2003anisotropic} and van der Waals ferromagnets\cite{ma2023anisotropic}. It is usually attributed to magnetic anisotropy\cite{bibes2000anisotropic,bibes2005anisotropic}, relaxation time anisotropy\cite{dai2022fourfold}, spin-dependent scattering near antiphase boundary\cite{eerenstein2002spin,ando2003anisotropic} and lattice symmetry\cite{ramos2008anomalous}.

Anisotropic magnetoresistance of FePd$_2$Te$_2$ in Fig. 2(b)-(j) was collected under field as high as 9 T. Such a high field is enough to align the magnetic moments along the direction of the field and subsequently exclude the possibility of magnetic anisotropy. This is also suggested by the fact that the peak of anisotropic magnetoresistance occurs along the direction $H \parallel I$ rather than $M \parallel I$. Aligned magnetic moments make direction of $H$ and $M$ equivalent. Considering the twinning effect, the high-order anisotropic magnetoresistance should be closely related to the pseudo four-fold lattice symmetry induced by perpendicular Fe chains. Twin boundary is also needed to be considered. In epitaxial Fe$_3$O$_4$ films\cite{li2010origin,margulies1997origin} and high $T_c$ cuprates\cite{ando2003anisotropic}, four-fold anisotropic magnetoresistance is attributed to antiphase antiferromagnetic domain boundaries\cite{eerenstein2002spin}. This physical picture of spin-polarized transport across sharp antiferromagnetic boundaries also applies to FePd$_2$Te$_2$. In FePd$_2$Te$_2$, magnetic moments near crystal domain boundary of orthorhombic twins with in-plane anisotropic Fe chains can be naturally resolved ferromagnetic and antiferromagnetic components at zero field\cite{shi2024fepd2te2}. At low fields, moments in discrete twins are gradually aligned along the direction of $H$. At high field, the moments near twin boundary components coupled with antiferromagnetism start to align along $H$. This also leads to additional high-order contribution in anisotropic magnetoresistance. High-field four-fold anisotropic magnetoresistance in FePd$_2$Te$_2$ is quite similar to Fe$_3$O$_4$\cite{li2010origin}, and is opposite to manganite films\cite{bibes2000anisotropic,bibes2005anisotropic}. Low-field anisotropic magnetoresistance is still needed for further analysis. Also, spin-polarized transport across sharp antiferromagnetic boundaries results to non-saturating magnetization\cite{margulies1997origin} and linear magnetoresistance\cite{eerenstein2002spin}, which has been observed in FePd$_2$Te$_2$ and evidenced by critical exponents obtained below. Since four-fold anisotropic magnetoresistance corresponds to pseudo four-fold symmetry from perpendicular Fe chains and resulting atomic sharp twin boundary with partially antiferromagnetic coupling, the additional four-fold anisotropic magnetoresistance should be able to be modulated by the fraction of twin boundaries. Figure 2(k) shows the value of two sets of anisotropic magnetoresistance peaks of FePd$_2$Te$_2$ and FePd$_{2.08}$Te$_2$, respectively. Strain from 4\% Pd atoms influence the crystal twins and subsequently changes anisotropic magnetoresistance successfully.

\begin{figure}
    \centering
    \includegraphics[width=1.0\linewidth]{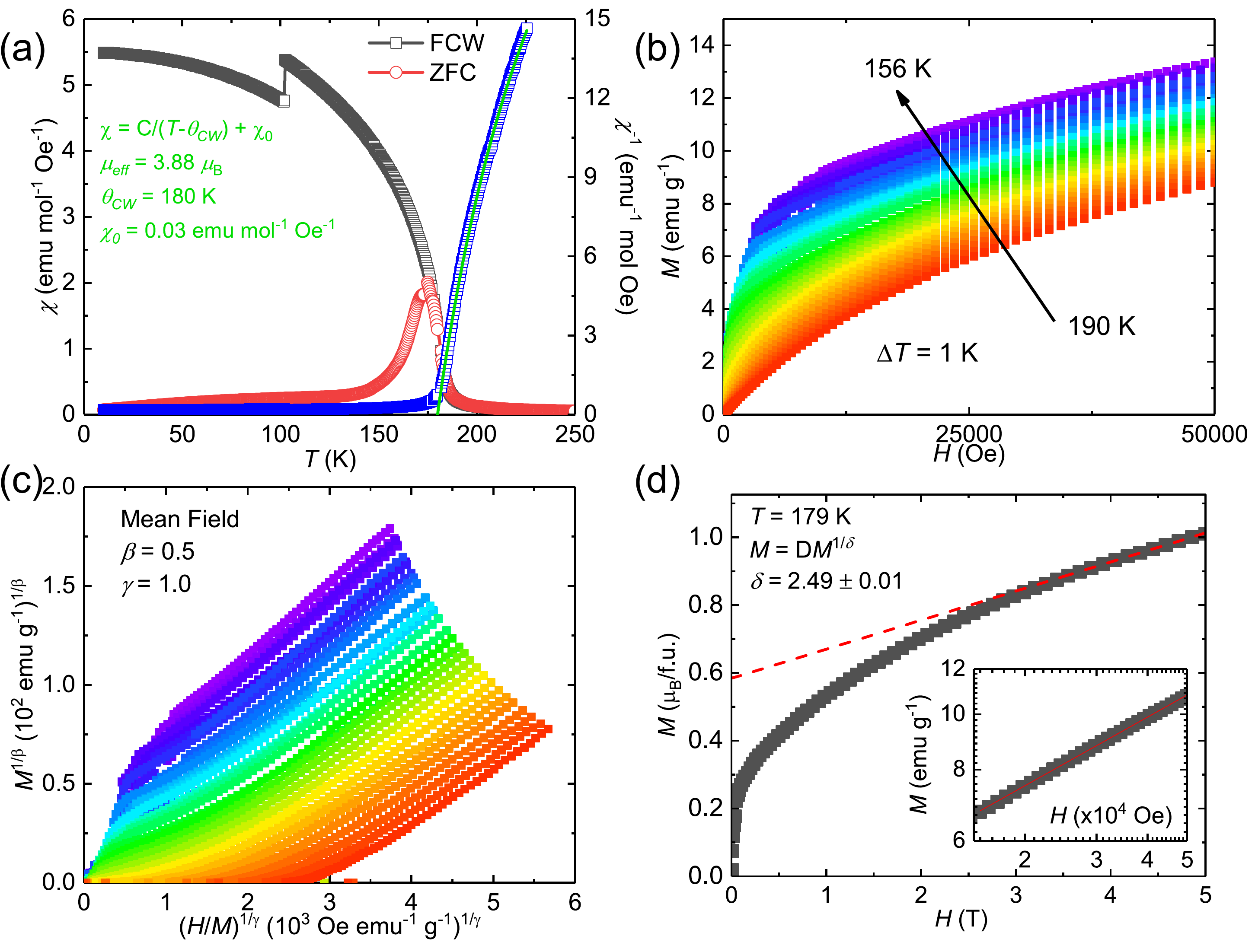}
    \caption{Magnetic Properties and entropy change. (a) Temperature-dependent magnetic susceptibility and reverse magnetic susceptibility along $H$//$c$-axis. Black, red and blue lines represent zero-field-cooling, field-cooling-warming magnetic susceptibility and reverse magnetic susceptibility, respectively. The anomaly around 100 K is a break-point due to measurement error. (b) Initial isothermal magnetization along $H$//$c$-axis near Curie temperature. (c) Arrott plot of $M^2$ vs $H/M$. (d) Initial isothermal magnetization at 179 K for FePd$_2$Te$_2$ along $H$//$c$-axis. The saturated magnetic moment is 0.58 $\mu_B$. Inset is log-log scale with a fitting curve.}
    \label{Fig. 3}
\end{figure}
Besides the influence of crystal twins on electrical transport, detailed measurements were performed to investigate the effect on magnetic properties in FePd$_2$Te$_2$. Figure 3(a) shows the magnetic susceptibility under 1000 Oe with  $H$//$c$-axis. A ferromagnetic-paramagnetic magnetic transition occurs around 180 K. Curie-Weiss fitting gives a effective moment $\mu_{eff}$ with 3.9 $\mu_B$. Besides, a downturn was observed in $\chi^{-1}$ at around 200 K, corresponding to the starting temperature of anisotropic magnetoresistance. A Hopkinson peak was observed just below $T_c$. Such obvious Hopkinson peak is rather rare in other van der Waals ferromagnets, so twinning effect may cause more complex magnetic domains in FePd$_2$Te$_2$.

Initial isothermal magnetization along $c$-axis near $T_c$ was measured and shown in Fig. 3(b). Arrott-Stokes equations are usually employed to analyze the critical behaviors of a magnetic phase transition as below\cite{arrott1967approximate},
\begin{equation}
    (\frac{H}{M})^{1/\gamma}=a\varepsilon+bM^{1/\beta},
\end{equation}
where $\varepsilon$ is reduced temperature $\varepsilon=(T-T_c)/T_c$, and $a$ and $b$ are constants. Six different kinds of theoretical models with short-range interactions were used to construct the modified Arrott plots, which are mean field model ($\beta$=0.5, $\gamma$=1.0), tricritical mean field model ($\beta$=0.25, $\gamma$=1.0), 3D Heisenberg model ($\beta$=0.365, $\gamma$=1.386), 3D XY model ($\beta$=0.345, $\gamma$=1.316), 3D Ising model ($\beta$=0.325, $\gamma$=1.24), 2D Ising model ($\beta$=0.125, $\gamma$=1.75). The mean field model was shown in Fig. 3(c), while the others in Supporting Information. In general, the normalized slop is a criterion to judge whether the model is appropriate. It is defined as slope($T$)/slope($T_c$), where slope($T$)=d$M^{1/\beta}$/d($H$/$M$)$^{1/\gamma}$. If the theoretical model is appropriate, the temperature-dependent normalized slope should be equal to $1$. Figure S1 gave the temperature dependence of normalized slope. Both above $T_c$ and below $T_c$, mean field model was the most appropriate. However, it was found that $T_c$ obtained from conventional Arrott plot did not match real $T_c$ well, which suggest mean field model was not accurate enough. 

Firstly, $\delta$ was fitted from equation,
\begin{equation}
    M=DH^{1/\delta}, T=T_c,
\end{equation}
where D is critical amplitudes\cite{fisher1967theory}. Figure 3(d) shows the initial isothermal magnetization at $T_c$. $\delta$ was obtained as 2.49$\pm$0.01 from the log-log scale in the inset of Fig. 3 (d), i.e., ln$M$=1/$\delta$ln$M$+ln$D$.

\begin{figure}
    \centering
    \includegraphics[width=1.0\linewidth]{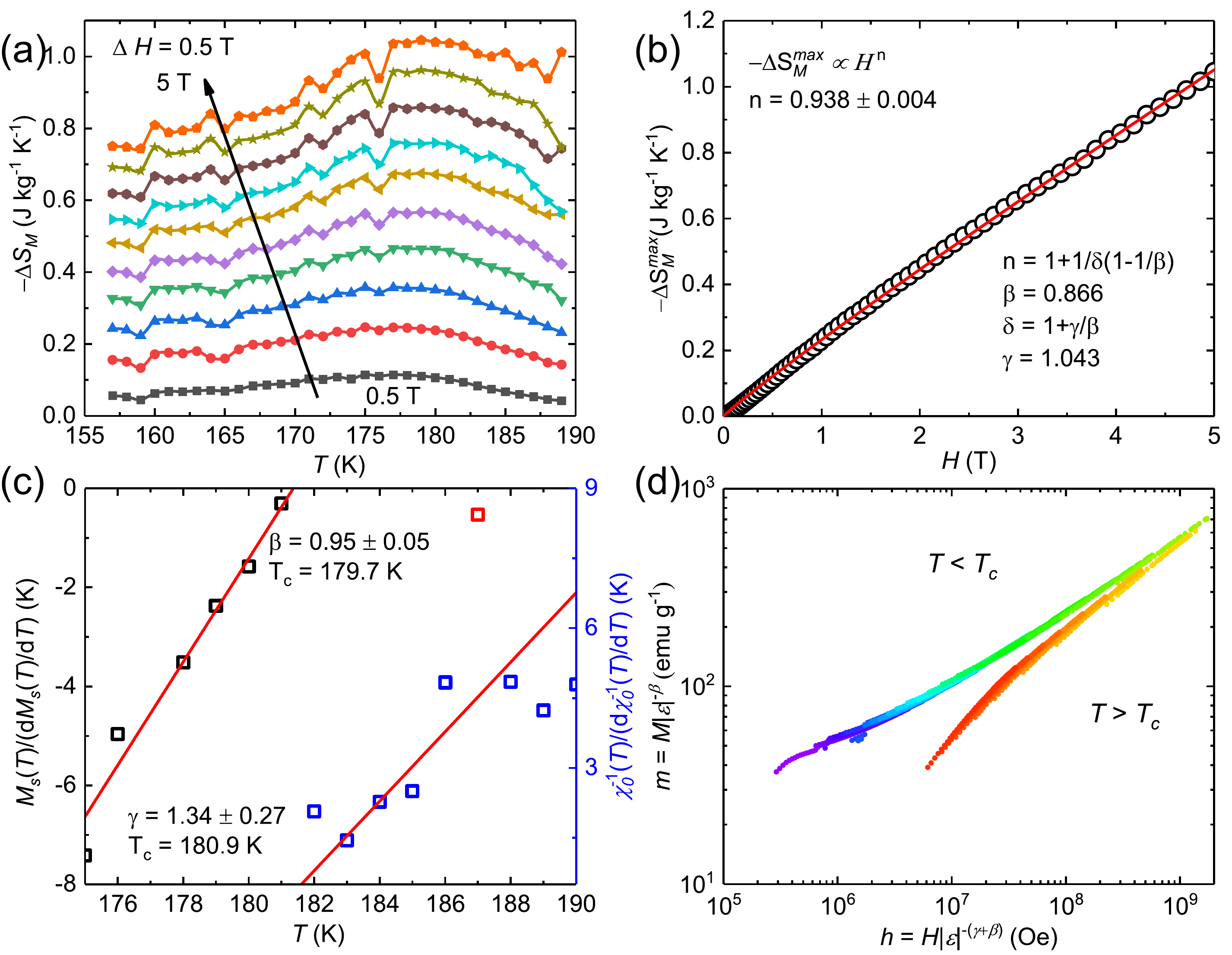}
    \caption{Critical behavior and scaling analysis.  (a) The magnetic entropy change $-\Delta S_M$ obtained from initial isothermal magnetization with $H$//$c$-axis. (b) Field-dependence of the maximum magnetic entropy change $-\Delta S_M^{max}$. (c) Kouvel-Fisher plots, termperature dependence of $M_S$(d$M_S$/d$T$)$^{-1}$ (left axis) and $\chi_0^{-1}$(d$\chi_0^{-1}$/d$T$)$^{-1}$. Solid line represents fitting curve. (d) Scaling plot of renormalized magnetization $m$ vs renormalized magnetic field $h$ in log-log scale below and above $T_c$ at high-field region.}
    \label{Fig. 4}
\end{figure}

\begin{table*}[htbp]
\caption{\label{tab:fonts}Critical exponents of FePd$_2$Te$_2$ and different theoretical models}
\begin{ruledtabular}
\begin{tabular}{cccccc}
Composition  & Technique & $\beta$ & $\gamma$ & $\delta$\\
\hline
FePd$_2$Te$_2$  & Critical isotherm &   &   & 2.49(1)\\
  & Magnetic entropy analysis & 0.87 & 1.04 & 2.20\\
  & Kouvel-Fisher method & 0.95(5) & 1.34(3) & 2.41(2)\\
Mean field & Theory & 0.5 & 1.0 & 3.0\\
Tricritical mean field  & Theory & 0.25 & 1.0 & 5\\
3D Heisenberg  & Theory & 0.365 & 1.386 & 4.8\\
3D XY  & Theory & 0.345 & 1.316 & 4.81\\
3D Ising  & Theory & 0.325 & 1.24 & 4.82\\
2D Ising  & Theory & 0.125 & 1.75 & 15\\
\end{tabular}
\end{ruledtabular}
\end{table*}

The magnetic entropy change also suggests the critical exponents of magnetic phase transition, which is closely related to the intrinsic magnetic coupling in FePd$_2$Te$_2$. Based Maxwell's relation, the magnetic entropy change is 
\begin{equation}
    \Delta S_M(T,H)=\int^H_0(\frac{\partial S}{\partial H})_TdH=\int^H_0(\frac{\partial M}{\partial T})_HdH.
\end{equation}
According to above equation, Fig. 4(a) showed the calculated magnetic entropy change $-\Delta S_M(T,H)$ as a function of temperature under several magnetic fields along $c$-axis. All $-\Delta S_M(T,H)$ curves exhibit a peak at $T_c$. The maxima negative magnetic entropy change $-\Delta S_M(T,H)$ reached 1 J kg$^{-1}$ K$^{-1}$, which is at the same order of magnitude as other van der Waals magnets\cite{wang2022magnetic}. In Fig. 4(b), field-dependent $-\Delta S_M^{max}$ was fitted by
\begin{equation}
    -\Delta S_M^{max}=aH^n
\end{equation}
The fitting result gave $n$=0.938(4). There is relation among $n$, $\delta$, $\beta$, $\gamma$,
\begin{gather}
    n=1+\frac{1}{\delta}(1-\frac{1}{\beta}),\\
    \delta=1+\frac{\gamma}{\beta},
\end{gather}
Combining $\delta$ = 2.49 $\pm$ 0.01, $\beta$ and $\gamma$ was calculated as 0.866 and 1.043, respectively. New $\delta$ was calculated as 2.20 through Eq. (6).

We tried to get precise critical exponents through iterative method of modified Arrott plot\cite{pramanik2009critical}, but it always diverged and got a $T_c$ beyond the temperature range measured. So, similar values from magnetic entropy analysis were employed to construct modified Arrott plot $M^{1/\beta}$ vs $(H/M)^{1/\gamma}$ with $\beta$ = 0.87 and $\gamma$ = 1.29. Spontaneous magnetization $M_S$ and inverse magnetic susceptibility $\chi_0^{-1}$ are obtained from the intercepts with the axis $M^{1/\beta}$ and $(H/M)^{1/\gamma}$ in the constructed modified Arrott plot. Through the temperature-dependent $M_S$ and $\chi_0^{-1}$, the Kouvel-Fisher method is used to determined the critical exponents\cite{kouvel1964detailed},
\begin{gather}
    \frac{M_S(T)}{dM_S(T)/dT}=\frac{T-T_c}{\beta},\\
    \frac{\chi_0^{-1}(T)}{d\chi_0^{-1}(T)/dT}=\frac{T-T_c}{\gamma}.
\end{gather}
The results are shown in Fig. 4(c), giving $\beta$ = 0.95 $\pm$ 0.05 with $T_c$ = 179.7 K and $\gamma$ = 1.34 $\pm$ 0.27 with $T_c$ = 180.9 K, respectively. These results are close to the value calculated from magnetic entropy change analysis. Another samples were used to analyze the critical exponents, and the critical exponents obtained in these two processes are similar.

Besides cross validation by isothermal magnetization analysis at $T_c$, magnetic entropy change analysis and Kouvel-Fisher method, scaling analysis is also used to identify the reliability of critical exponents. The rescaled magnetization and rescaled magnetic field are defined as 
\begin{gather}
    m=\varepsilon^{-\beta}M(H,\varepsilon),\\\
    h=\varepsilon^{-(\beta+\gamma)}H,
\end{gather}
where $\varepsilon$ is reduced temperature. The magnetic equation of state in the critical region is
\begin{equation}
    m=f_\pm (h).
\end{equation}
This means that with right critical exponents, $h$-dependent $m$ will be collapsed into two curves above and below $T_c$, respectively. In Fig. 4(d), two branches are observed clearly, which verifies the reliability of the obtained critical exponents.

Table I lists critical exponents of FePd$_2$Te$_2$ and different theoretical models with short-range interactions. Due to relatively large error bar in critical exponents obtained through Kouvel-Fisher method, the following discussions are based on those obtained by magnetic entropy analysis. In order to clarify the relation between crystal structure and magnetic coupling, we try to solve the space dimensionality $d$ and spin dimensionality $n$ of FePd$_2$Te$_2$. According to renormalization group analysis\cite{fisher1972critical}, the interactions between spins decay with spin-spin distance $r$ as $J(r)\sim r^{-(d+\sigma)}$, where $d$ is the space dimensionality and $\sigma$ is constant. $\sigma > $ 2 suggests a Heisenberg-type short-range interactions, corresponding to the theoretical models in Table I. It is obvious that magnetic coupling in FePd$_2$Te$_2$  cannot be described by any model with short-range interactions, i.e., $\sigma >$ 2. 

When $\sigma$ lies between $d/2$ and 2, the interactions cannot be described by single type short- or long-range model. This also induces a continuous change of critical exponent 
 $\gamma$. Almost all van der Waals magnets are located in this region\cite{wang2022magnetic}. In this case, the critical exponent $\gamma$ is predicted as\cite{fisher1972critical}
\begin{equation}
\begin{aligned}
    \gamma=&1+\frac{4}{d}\frac{n+2}{n+8}\Delta\sigma+\frac{8(n+2)(n-4)}{d^2(n+8)^2}\\&\times(1+\frac{2G(d/2)(7n+20)}{(n-4)(n+8)})\Delta\sigma^2,
\end{aligned}
\end{equation}
where 
\begin{align}
    \Delta\sigma&=(\sigma-d/2),\\
    G(d/2)&=3-1/4\times(d/2)^2.
\end{align}
We adjusted $\sigma$ and different sets of \{$d$:$n$\} to yield a value for $\gamma$ close to that experimentally observed, i.e., $\gamma$ = 1.043. The obtained $\sigma$ can be use to calculate other critical exponents through following equations,
\begin{align}
    v=&\gamma/\sigma,\\
    \alpha=&2-v d,\\
    \beta=&(2-\alpha-\gamma)/2,\\
    \delta=&1+\gamma/\beta.
\end{align}
However, with $\gamma$ closest to that experimentally observed, $\beta$ and $\delta$ do not match corresponding experimental values. 

These results inspired us to reconsider the magnetic coupling from a perspective of pure long-range interactions, i.e., $\Delta \sigma = \sigma -d/2<0$. In this regime, the exponents are calculated as below\cite{fisher1972critical},
\begin{align}
    \gamma = &1,\\
    \eta =& 2-\sigma,\\
    v =&1/\sigma.
\end{align}
However, renormalization group analysis argues that the critical exponents take classical mean-field values in this region\cite{fisher1972critical}. Although $\gamma$ approaches 1, $\beta$ is much larger than that in mean field model. Interestingly, the critical exponents of FePd$_2$Te$_2$ do not correspond to any conventional model, regardless of with long- or short-range interactions. A possible reason is that FePd$_2$Te$_2$ possesses different critical behavior below and above $T_c$, like the case in Pr$_{0.6-x}$Er$_x$Ca$_{0.1}$Sr$_{0.3}$MnO$_3$\cite{omrani2016critical}, Cu$_2$OSeO$_3$\cite{chauhan2022different} and FeGe\cite{zhang2016critical}. The difference is that $\beta$ is not close to any universal class.

To our best knowledge, the critical exponents of FePd$_2$Te$_2$ are the first set of values beyond conventional models in those of van der Waals magnets, which suggest the unconventional magnetic interactions and coupling. Due to 
\begin{align}
    M_S (T) = &M_0(-\varepsilon)^\beta, \varepsilon<0, T<T_c,\\
    \chi_0^{-1}(T)=&(h_0/m_0)\varepsilon^\gamma, \varepsilon>0, T>T_c,
\end{align}
$\gamma$ and $\beta$ reflect the divergence of inverse magnetic susceptibility above $T_c$ and the growth speed of spontaneous magnetization below $T_c$, respectively. In FePd$_2$Te$_2$, the critical exponent $\gamma$ close to 1 suggests a mean field model with isotropic long-range interactions above $T_c$. $\beta$ with value 0.87 deviates from the prediction of the mean field model and is rather large. Such large $\beta$ is never estimated in other van der Waals magnets, whose $\beta$ is usually around 0.3\cite{wang2022magnetic}. Compared to those of other van der Waals magnets, $\delta$ of FePd$_2$Te$_2$ is also small. Such unconventional large $\beta$ and small $\gamma$ is also observed in doped manganites such as Ca$_{0.85}$Sm$_{0.15}$MnO$_3$\cite{nag2019magnetic}, Ca$_{0.85}$Dy$_{0.15}$MnO$_3$\cite{sarkar2021manifestation} and Nd$_{0.4}$Sr$_{0.6}$MnO$_3$ nanoparticles\cite{kundu2013critical}. 

Large value of $\beta$ means slow growth of spontaneous magnetization, while small $\delta$ value results from the non-saturating behavior of field-dependent magnetization. The slow growth of spontaneous magnetization in FePd$_2$Te$_2$ can be attributed to measurement along $c$-axis, which is hard axis. Similar case occurs in Fe$_{2.72}$GeTe$_2$ with perpendicular magnetic anisotropy. In Fe$_{2.72}$GeTe$_2$, $\beta$ along $c$-axis is 0.361(3), while 0.714(3) within $ab$-plane\cite{liu2019field}. As for small $\delta$, non-saturating behavior in other materials is usually attributed to an interface magnetism resulting in a current of spin-polarized electrons\cite{rumpf2010non}, little percent of antiferromagnetic phase\cite{sarkar2021manifestation}, size-dependent ferromagnetism\cite{kundu2013critical} and coexistence of non-ferromagnetic phases along with percolating ferromagnetic-clusters\cite{kumar2012critical}. Also, finite-size ferromagnetic clusters in paramagnetic phases also leads to Griffiths phase, which is also one of the reasons for small $\gamma$\cite{nag2019magnetic}. All these factors involve nano-structure in magnetic systems. In FePd$_2$Te$_2$, atomic twin boundary and in-plane anisotropy perpendicular to Fe chains lead to a natural antiferromangetic component, which is responsible for non-saturating behavior. Fragmented Fe chains and magnetic clusters are also suggested. These results suggest the close relationship between the unusual critical exponents and structure in mesoscopic scale in FePd$_2$Te$_2$.

Although FePd$_2$Te$_2$ exhibits coercive field as high as 2.5 T, its $T_c$ is only $\sim$ 180 K, which is smaller than other van der Waals ferromagnets discovered recently, such as Fe$_3$GaTe$_2$\cite{zhang2022above}. In traditional theorem, finite-temperature long-range magnetic order is excluded in one-dimensional systems\cite{mermin1966absence}. Finite $T_c$ in FePd$_2$Te$_2$ can be attributed to two factors. One is that the Fe chains is nominally one-dimensional and actually kinked. The other is long-range magnetic interactions and interchain coupling. Thus, FePd$_2$Te$_2$ exhibits mean-field-like rather than one-dimensional ferromagnetism, although it crystallizes with quasi-one-dimensional Fe chains. In the same vein, enhancing interchain coupling is an effective way to increase $T_c$\cite{duan2022synthesis}. In metallic FePd$_2$Te$_2$, interchain coupling is mediated by itinerant electrons, so increasing carrier density through chemical doping in Pd/Te sites may be helpful. Electrostatic doping in devices based on thin flakes is not a good choice. Electrostatic doping usually requires small thickness down to few layers. However, the critical temperature $T_c$ in thin flakes with spin-spin correlation exceeding the thickness will shift to lower temperatures than that of bulk\cite{zhang2001thickness}. For FePd$_2$Te$_2$ with long-range spin interactions, this means that the critical thickness is larger than those of other van der Waals ferromagnets, whose critical thickness is usually $\sim$ 3 nm\cite{deng2018gate,chen2024high}.
\section{Summary}
In summary, influence of crystal twinning effect on electrical transport and magnetic properties are studied systemically. Easy-plane anisotropic orthorhombic crystal domains lead to natural antiferromagnetic coupling component near atomically sharp domain boundary and thus induce four-fold in-plane anisotropic magnetoresistance, which also reflects the pseudo four-fold lattice symmetry. The anisotropic magnetoresistance can be modulated by the fraction of domain boundary, thus by strain resulting from extra Pd atoms indirectly. Hopkinson effect observed in both Ac and DC magnetic susceptibility indicates intense magnetic domain motion, which is rather rare in other van der Waals ferromagnets. Unusual critical exponents $\beta = 0.866$, $\gamma = 1.043$, $\delta = 2.20$ do not belong to any conventional universal class and evidence a decoupling between one-dimensional Fe-chains and mean-field-like ferromagnetism with long-range interactions. Large $\beta$ and small $\gamma$ is closely related to antiferromagnetic coupling component near crystal domain walls. These results also suggest the possible presence of fragmented Fe chains and magnetic clusters in FePd$_2$Te$_2$. Twinning induced complicated spin texture, magnetic properties of detwinned FePd$_2$Te$_2$ and how to increase $T_c$ are still needed further studied.
\section{Acknowledge}
This work is supported by the National Natural Science Foundation of China (52250308) and the National Key Research and Development Program of China (2023YFA1406301). A portion of this work was carried out at the Synergetic Extreme Condition User Facility (SECUF).
\bibliography{apsguide4-2}

\providecommand{\noopsort}[1]{}\providecommand{\singleletter}[1]{#1}%
\begin{thebibliography}{58}%
\makeatletter
\providecommand \@ifxundefined [1]{%
 \@ifx{#1\undefined}
}%
\providecommand \@ifnum [1]{%
 \ifnum #1\expandafter \@firstoftwo
 \else \expandafter \@secondoftwo
 \fi
}%
\providecommand \@ifx [1]{%
 \ifx #1\expandafter \@firstoftwo
 \else \expandafter \@secondoftwo
 \fi
}%
\providecommand \natexlab [1]{#1}%
\providecommand \enquote  [1]{``#1''}%
\providecommand \bibnamefont  [1]{#1}%
\providecommand \bibfnamefont [1]{#1}%
\providecommand \citenamefont [1]{#1}%
\providecommand \href@noop [0]{\@secondoftwo}%
\providecommand \href [0]{\begingroup \@sanitize@url \@href}%
\providecommand \@href[1]{\@@startlink{#1}\@@href}%
\providecommand \@@href[1]{\endgroup#1\@@endlink}%
\providecommand \@sanitize@url [0]{\catcode `\\12\catcode `\$12\catcode `\&12\catcode `\#12\catcode `\^12\catcode `\_12\catcode `\%12\relax}%
\providecommand \@@startlink[1]{}%
\providecommand \@@endlink[0]{}%
\providecommand \url  [0]{\begingroup\@sanitize@url \@url }%
\providecommand \@url [1]{\endgroup\@href {#1}{\urlprefix }}%
\providecommand \urlprefix  [0]{URL }%
\providecommand \Eprint [0]{\href }%
\providecommand \doibase [0]{https://doi.org/}%
\providecommand \selectlanguage [0]{\@gobble}%
\providecommand \bibinfo  [0]{\@secondoftwo}%
\providecommand \bibfield  [0]{\@secondoftwo}%
\providecommand \translation [1]{[#1]}%
\providecommand \BibitemOpen [0]{}%
\providecommand \bibitemStop [0]{}%
\providecommand \bibitemNoStop [0]{.\EOS\space}%
\providecommand \EOS [0]{\spacefactor3000\relax}%
\providecommand \BibitemShut  [1]{\csname bibitem#1\endcsname}%
\let\auto@bib@innerbib\@empty
\bibitem [{\citenamefont {Parsons}(2003)}]{parsons2003introduction}%
  \BibitemOpen
  \bibfield  {author} {\bibinfo {author} {\bibfnamefont {S.}~\bibnamefont {Parsons}},\ }\bibfield  {title} {\bibinfo {title} {Introduction to {T}winning},\ }\href@noop {} {\bibfield  {journal} {\bibinfo  {journal} {Acta Crystallographica Section D: Biological Crystallography}\ }\textbf {\bibinfo {volume} {59}},\ \bibinfo {pages} {1995} (\bibinfo {year} {2003})}\BibitemShut {NoStop}%
\bibitem [{\citenamefont {Nespolo}(2015)}]{nespolo2015tips}%
  \BibitemOpen
  \bibfield  {author} {\bibinfo {author} {\bibfnamefont {M.}~\bibnamefont {Nespolo}},\ }\bibfield  {title} {\bibinfo {title} {Tips and {T}raps on {C}rystal {T}winning: {H}ow to {F}ully {D}escribe {Y}our {T}win},\ }\href@noop {} {\bibfield  {journal} {\bibinfo  {journal} {Crystal Research and Technology}\ }\textbf {\bibinfo {volume} {50}},\ \bibinfo {pages} {362} (\bibinfo {year} {2015})}\BibitemShut {NoStop}%
\bibitem [{\citenamefont {Fisher}\ \emph {et~al.}(2011)\citenamefont {Fisher}, \citenamefont {Degiorgi},\ and\ \citenamefont {Shen}}]{fisher2011plane}%
  \BibitemOpen
  \bibfield  {author} {\bibinfo {author} {\bibfnamefont {I.~R.}\ \bibnamefont {Fisher}}, \bibinfo {author} {\bibfnamefont {L.}~\bibnamefont {Degiorgi}},\ and\ \bibinfo {author} {\bibfnamefont {Z.}~\bibnamefont {Shen}},\ }\bibfield  {title} {\bibinfo {title} {In-plane {E}lectronic {A}nisotropy of {U}nderdoped ‘122’ {F}e-{A}rsenide {S}uperconductors {R}evealed by {M}easurements of {D}etwinned {S}ingle {C}rystals},\ }\href@noop {} {\bibfield  {journal} {\bibinfo  {journal} {Reports on progress in Physics}\ }\textbf {\bibinfo {volume} {74}},\ \bibinfo {pages} {124506} (\bibinfo {year} {2011})}\BibitemShut {NoStop}%
\bibitem [{\citenamefont {Specht}\ \emph {et~al.}(1988)\citenamefont {Specht}, \citenamefont {Sparks}, \citenamefont {Dhere}, \citenamefont {Brynestad}, \citenamefont {Cavin}, \citenamefont {Kroeger},\ and\ \citenamefont {Oye}}]{specht1988effect}%
  \BibitemOpen
  \bibfield  {author} {\bibinfo {author} {\bibfnamefont {E.}~\bibnamefont {Specht}}, \bibinfo {author} {\bibfnamefont {C.}~\bibnamefont {Sparks}}, \bibinfo {author} {\bibfnamefont {A.}~\bibnamefont {Dhere}}, \bibinfo {author} {\bibfnamefont {J.}~\bibnamefont {Brynestad}}, \bibinfo {author} {\bibfnamefont {O.}~\bibnamefont {Cavin}}, \bibinfo {author} {\bibfnamefont {D.}~\bibnamefont {Kroeger}},\ and\ \bibinfo {author} {\bibfnamefont {H.}~\bibnamefont {Oye}},\ }\bibfield  {title} {\bibinfo {title} {Effect of {O}xygen {P}ressure on the {O}rthorhombic-{T}etragonal {T}ransition in the {H}igh-{T}emperature {S}uperconductor {YB}a$_2${C}u$_3${O}$_x$},\ }\href@noop {} {\bibfield  {journal} {\bibinfo  {journal} {Physical Review B}\ }\textbf {\bibinfo {volume} {37}},\ \bibinfo {pages} {7426} (\bibinfo {year} {1988})}\BibitemShut {NoStop}%
\bibitem [{\citenamefont {Radaelli}\ \emph {et~al.}(1994)\citenamefont {Radaelli}, \citenamefont {Hinks}, \citenamefont {Mitchell}, \citenamefont {Hunter}, \citenamefont {Wagner}, \citenamefont {Dabrowski}, \citenamefont {Vandervoort}, \citenamefont {Viswanathan},\ and\ \citenamefont {Jorgensen}}]{radaelli1994structural}%
  \BibitemOpen
  \bibfield  {author} {\bibinfo {author} {\bibfnamefont {P.}~\bibnamefont {Radaelli}}, \bibinfo {author} {\bibfnamefont {D.}~\bibnamefont {Hinks}}, \bibinfo {author} {\bibfnamefont {A.}~\bibnamefont {Mitchell}}, \bibinfo {author} {\bibfnamefont {B.}~\bibnamefont {Hunter}}, \bibinfo {author} {\bibfnamefont {J.}~\bibnamefont {Wagner}}, \bibinfo {author} {\bibfnamefont {B.}~\bibnamefont {Dabrowski}}, \bibinfo {author} {\bibfnamefont {K.}~\bibnamefont {Vandervoort}}, \bibinfo {author} {\bibfnamefont {H.}~\bibnamefont {Viswanathan}},\ and\ \bibinfo {author} {\bibfnamefont {J.}~\bibnamefont {Jorgensen}},\ }\bibfield  {title} {\bibinfo {title} {Structural and {S}uperconducting {P}roperties of {L}a$_{2-x}${S}r$_x${C}u{O}$_4$ as a {F}unction of {S}r {c}ontent},\ }\href@noop {} {\bibfield  {journal} {\bibinfo  {journal} {Physical Review B}\ }\textbf {\bibinfo {volume} {49}},\ \bibinfo {pages} {4163} (\bibinfo {year} {1994})}\BibitemShut {NoStop}%
\bibitem [{\citenamefont {Tanatar}\ \emph {et~al.}(2009)\citenamefont {Tanatar}, \citenamefont {Kreyssig}, \citenamefont {Nandi}, \citenamefont {Ni}, \citenamefont {Bud’ko}, \citenamefont {Canfield}, \citenamefont {Goldman},\ and\ \citenamefont {Prozorov}}]{tanatar2009direct}%
  \BibitemOpen
  \bibfield  {author} {\bibinfo {author} {\bibfnamefont {M.}~\bibnamefont {Tanatar}}, \bibinfo {author} {\bibfnamefont {A.}~\bibnamefont {Kreyssig}}, \bibinfo {author} {\bibfnamefont {S.}~\bibnamefont {Nandi}}, \bibinfo {author} {\bibfnamefont {N.}~\bibnamefont {Ni}}, \bibinfo {author} {\bibfnamefont {S.~L.}\ \bibnamefont {Bud’ko}}, \bibinfo {author} {\bibfnamefont {P.}~\bibnamefont {Canfield}}, \bibinfo {author} {\bibfnamefont {A.}~\bibnamefont {Goldman}},\ and\ \bibinfo {author} {\bibfnamefont {R.}~\bibnamefont {Prozorov}},\ }\bibfield  {title} {\bibinfo {title} {Direct {I}maging of the {S}tructural {D}omains in the {I}ron {P}nictides {AF}e$_2${A}s$_2$ ({A}= {C}a, {S}r, {B}a)},\ }\href@noop {} {\bibfield  {journal} {\bibinfo  {journal} {Physical Review B—Condensed Matter and Materials Physics}\ }\textbf {\bibinfo {volume} {79}},\ \bibinfo {pages} {180508} (\bibinfo {year} {2009})}\BibitemShut {NoStop}%
\bibitem [{\citenamefont {Prozorov}\ \emph {et~al.}(2009)\citenamefont {Prozorov}, \citenamefont {Tanatar}, \citenamefont {Ni}, \citenamefont {Kreyssig}, \citenamefont {Nandi}, \citenamefont {Bud’Ko}, \citenamefont {Goldman},\ and\ \citenamefont {Canfield}}]{prozorov2009intrinsic}%
  \BibitemOpen
  \bibfield  {author} {\bibinfo {author} {\bibfnamefont {R.}~\bibnamefont {Prozorov}}, \bibinfo {author} {\bibfnamefont {M.}~\bibnamefont {Tanatar}}, \bibinfo {author} {\bibfnamefont {N.}~\bibnamefont {Ni}}, \bibinfo {author} {\bibfnamefont {A.}~\bibnamefont {Kreyssig}}, \bibinfo {author} {\bibfnamefont {S.}~\bibnamefont {Nandi}}, \bibinfo {author} {\bibfnamefont {S.}~\bibnamefont {Bud’Ko}}, \bibinfo {author} {\bibfnamefont {A.}~\bibnamefont {Goldman}},\ and\ \bibinfo {author} {\bibfnamefont {P.}~\bibnamefont {Canfield}},\ }\bibfield  {title} {\bibinfo {title} {Intrinsic {P}inning on {S}tructural {D}omains in {U}nderdoped {S}ingle {C}rystals of {B}a({F}e$_{1-x}${C}o$_x$)$_2${A}s$_2$},\ }\href@noop {} {\bibfield  {journal} {\bibinfo  {journal} {Physical Review B—Condensed Matter and Materials Physics}\ }\textbf {\bibinfo {volume} {80}},\ \bibinfo {pages} {174517} (\bibinfo {year} {2009})}\BibitemShut {NoStop}%
\bibitem [{\citenamefont {Li}\ \emph {et~al.}(2009)\citenamefont {Li}, \citenamefont {de~La~Cruz}, \citenamefont {Huang}, \citenamefont {Chen}, \citenamefont {Xia}, \citenamefont {Luo}, \citenamefont {Wang},\ and\ \citenamefont {Dai}}]{li2009structural}%
  \BibitemOpen
  \bibfield  {author} {\bibinfo {author} {\bibfnamefont {S.}~\bibnamefont {Li}}, \bibinfo {author} {\bibfnamefont {C.}~\bibnamefont {de~La~Cruz}}, \bibinfo {author} {\bibfnamefont {Q.}~\bibnamefont {Huang}}, \bibinfo {author} {\bibfnamefont {G.}~\bibnamefont {Chen}}, \bibinfo {author} {\bibfnamefont {T.-L.}\ \bibnamefont {Xia}}, \bibinfo {author} {\bibfnamefont {J.}~\bibnamefont {Luo}}, \bibinfo {author} {\bibfnamefont {N.}~\bibnamefont {Wang}},\ and\ \bibinfo {author} {\bibfnamefont {P.}~\bibnamefont {Dai}},\ }\bibfield  {title} {\bibinfo {title} {Structural and {M}agnetic {P}hase {T}ransitions in {N}a$_{1-\delta}${F}e{A}s},\ }\href@noop {} {\bibfield  {journal} {\bibinfo  {journal} {Physical Review B—Condensed Matter and Materials Physics}\ }\textbf {\bibinfo {volume} {80}},\ \bibinfo {pages} {020504} (\bibinfo {year} {2009})}\BibitemShut {NoStop}%
\bibitem [{\citenamefont {McQueen}\ \emph {et~al.}(2009)\citenamefont {McQueen}, \citenamefont {Williams}, \citenamefont {Stephens}, \citenamefont {Tao}, \citenamefont {Zhu}, \citenamefont {Ksenofontov}, \citenamefont {Casper}, \citenamefont {Felser},\ and\ \citenamefont {Cava}}]{mcqueen2009tetragonal}%
  \BibitemOpen
  \bibfield  {author} {\bibinfo {author} {\bibfnamefont {T.}~\bibnamefont {McQueen}}, \bibinfo {author} {\bibfnamefont {A.}~\bibnamefont {Williams}}, \bibinfo {author} {\bibfnamefont {P.}~\bibnamefont {Stephens}}, \bibinfo {author} {\bibfnamefont {J.}~\bibnamefont {Tao}}, \bibinfo {author} {\bibfnamefont {Y.}~\bibnamefont {Zhu}}, \bibinfo {author} {\bibfnamefont {V.}~\bibnamefont {Ksenofontov}}, \bibinfo {author} {\bibfnamefont {F.}~\bibnamefont {Casper}}, \bibinfo {author} {\bibfnamefont {C.}~\bibnamefont {Felser}},\ and\ \bibinfo {author} {\bibfnamefont {R.}~\bibnamefont {Cava}},\ }\bibfield  {title} {\bibinfo {title} {Tetragonal-to-{O}rthorhombic {S}tructural {P}hase {T}ransition at 90 {K} in the {S}uperconductor {F}e$_{1.01}${S}e},\ }\href@noop {} {\bibfield  {journal} {\bibinfo  {journal} {Physical Review Letters}\ }\textbf {\bibinfo {volume} {103}},\ \bibinfo {pages} {057002} (\bibinfo {year} {2009})}\BibitemShut {NoStop}%
\bibitem [{\citenamefont {Dolan}\ \emph {et~al.}(1989)\citenamefont {Dolan}, \citenamefont {Chandrashekhar}, \citenamefont {Dinger}, \citenamefont {Feild},\ and\ \citenamefont {Holtzberg}}]{dolan1989vortex}%
  \BibitemOpen
  \bibfield  {author} {\bibinfo {author} {\bibfnamefont {G.}~\bibnamefont {Dolan}}, \bibinfo {author} {\bibfnamefont {G.}~\bibnamefont {Chandrashekhar}}, \bibinfo {author} {\bibfnamefont {T.}~\bibnamefont {Dinger}}, \bibinfo {author} {\bibfnamefont {C.}~\bibnamefont {Feild}},\ and\ \bibinfo {author} {\bibfnamefont {F.}~\bibnamefont {Holtzberg}},\ }\bibfield  {title} {\bibinfo {title} {Vortex {S}tructure in {YB}a$_2${C}u$_3${O}$_7$ and {E}vidence for {I}ntrinsic {P}inning},\ }\href@noop {} {\bibfield  {journal} {\bibinfo  {journal} {Physical review letters}\ }\textbf {\bibinfo {volume} {62}},\ \bibinfo {pages} {827} (\bibinfo {year} {1989})}\BibitemShut {NoStop}%
\bibitem [{\citenamefont {Maggio-Aprile}\ \emph {et~al.}(1997)\citenamefont {Maggio-Aprile}, \citenamefont {Renner}, \citenamefont {Erb}, \citenamefont {Walker},\ and\ \citenamefont {Fischer}}]{maggio1997critical}%
  \BibitemOpen
  \bibfield  {author} {\bibinfo {author} {\bibfnamefont {I.}~\bibnamefont {Maggio-Aprile}}, \bibinfo {author} {\bibfnamefont {C.}~\bibnamefont {Renner}}, \bibinfo {author} {\bibfnamefont {A.}~\bibnamefont {Erb}}, \bibinfo {author} {\bibfnamefont {E.}~\bibnamefont {Walker}},\ and\ \bibinfo {author} {\bibfnamefont {{\O}.}~\bibnamefont {Fischer}},\ }\bibfield  {title} {\bibinfo {title} {Critical {C}urrents {A}pproaching the {D}epairing {L}imit at a {T}win {B}oundary in {YB}a$_2${C}u$_3${O}$_{7-\delta}$},\ }\href@noop {} {\bibfield  {journal} {\bibinfo  {journal} {Nature}\ }\textbf {\bibinfo {volume} {390}},\ \bibinfo {pages} {487} (\bibinfo {year} {1997})}\BibitemShut {NoStop}%
\bibitem [{\citenamefont {Song}\ \emph {et~al.}(2012)\citenamefont {Song}, \citenamefont {Wang}, \citenamefont {Jiang}, \citenamefont {Wang}, \citenamefont {He}, \citenamefont {Chen}, \citenamefont {Hoffman}, \citenamefont {Ma},\ and\ \citenamefont {Xue}}]{song2012suppression}%
  \BibitemOpen
  \bibfield  {author} {\bibinfo {author} {\bibfnamefont {C.-L.}\ \bibnamefont {Song}}, \bibinfo {author} {\bibfnamefont {Y.-L.}\ \bibnamefont {Wang}}, \bibinfo {author} {\bibfnamefont {Y.-P.}\ \bibnamefont {Jiang}}, \bibinfo {author} {\bibfnamefont {L.}~\bibnamefont {Wang}}, \bibinfo {author} {\bibfnamefont {K.}~\bibnamefont {He}}, \bibinfo {author} {\bibfnamefont {X.}~\bibnamefont {Chen}}, \bibinfo {author} {\bibfnamefont {J.~E.}\ \bibnamefont {Hoffman}}, \bibinfo {author} {\bibfnamefont {X.-C.}\ \bibnamefont {Ma}},\ and\ \bibinfo {author} {\bibfnamefont {Q.-K.}\ \bibnamefont {Xue}},\ }\bibfield  {title} {\bibinfo {title} {Suppression of {S}uperconductivity by {T}win {B}oundaries in {F}e{S}e},\ }\href@noop {} {\bibfield  {journal} {\bibinfo  {journal} {Physical Review Letters}\ }\textbf {\bibinfo {volume} {109}},\ \bibinfo {pages} {137004} (\bibinfo {year} {2012})}\BibitemShut {NoStop}%
\bibitem [{\citenamefont {Tsuei}\ and\ \citenamefont {Kirtley}(2000)}]{tsuei2000pairing}%
  \BibitemOpen
  \bibfield  {author} {\bibinfo {author} {\bibfnamefont {C.}~\bibnamefont {Tsuei}}\ and\ \bibinfo {author} {\bibfnamefont {J.}~\bibnamefont {Kirtley}},\ }\bibfield  {title} {\bibinfo {title} {Pairing {S}ymmetry in {C}uprate {S}uperconductors},\ }\href@noop {} {\bibfield  {journal} {\bibinfo  {journal} {Reviews of Modern Physics}\ }\textbf {\bibinfo {volume} {72}},\ \bibinfo {pages} {969} (\bibinfo {year} {2000})}\BibitemShut {NoStop}%
\bibitem [{\citenamefont {Watashige}\ \emph {et~al.}(2015)\citenamefont {Watashige}, \citenamefont {Tsutsumi}, \citenamefont {Hanaguri}, \citenamefont {Kohsaka}, \citenamefont {Kasahara}, \citenamefont {Furusaki}, \citenamefont {Sigrist}, \citenamefont {Meingast}, \citenamefont {Wolf}, \citenamefont {L{\"o}hneysen} \emph {et~al.}}]{watashige2015evidence}%
  \BibitemOpen
  \bibfield  {author} {\bibinfo {author} {\bibfnamefont {T.}~\bibnamefont {Watashige}}, \bibinfo {author} {\bibfnamefont {Y.}~\bibnamefont {Tsutsumi}}, \bibinfo {author} {\bibfnamefont {T.}~\bibnamefont {Hanaguri}}, \bibinfo {author} {\bibfnamefont {Y.}~\bibnamefont {Kohsaka}}, \bibinfo {author} {\bibfnamefont {S.}~\bibnamefont {Kasahara}}, \bibinfo {author} {\bibfnamefont {A.}~\bibnamefont {Furusaki}}, \bibinfo {author} {\bibfnamefont {M.}~\bibnamefont {Sigrist}}, \bibinfo {author} {\bibfnamefont {C.}~\bibnamefont {Meingast}}, \bibinfo {author} {\bibfnamefont {T.}~\bibnamefont {Wolf}}, \bibinfo {author} {\bibfnamefont {H.~v.}\ \bibnamefont {L{\"o}hneysen}}, \emph {et~al.},\ }\bibfield  {title} {\bibinfo {title} {Evidence for {T}ime-{R}eversal {S}ymmetry {B}reaking of the {S}uperconducting {S}tate near {T}win-{B}oundary {I}nterfaces in {F}e{S}e {R}evealed by {S}canning {T}unneling {S}pectroscopy},\ }\href@noop {} {\bibfield  {journal} {\bibinfo  {journal} {Physical Review X}\ }\textbf {\bibinfo {volume}
  {5}},\ \bibinfo {pages} {031022} (\bibinfo {year} {2015})}\BibitemShut {NoStop}%
\bibitem [{\citenamefont {Shi}\ \emph {et~al.}(2024)\citenamefont {Shi}, \citenamefont {Geng}, \citenamefont {Wang}, \citenamefont {Yang}, \citenamefont {Shang}, \citenamefont {Wang}, \citenamefont {Mi}, \citenamefont {Huang}, \citenamefont {Pan}, \citenamefont {Gui} \emph {et~al.}}]{shi2024fepd2te2}%
  \BibitemOpen
  \bibfield  {author} {\bibinfo {author} {\bibfnamefont {B.}~\bibnamefont {Shi}}, \bibinfo {author} {\bibfnamefont {Y.}~\bibnamefont {Geng}}, \bibinfo {author} {\bibfnamefont {H.}~\bibnamefont {Wang}}, \bibinfo {author} {\bibfnamefont {J.}~\bibnamefont {Yang}}, \bibinfo {author} {\bibfnamefont {C.}~\bibnamefont {Shang}}, \bibinfo {author} {\bibfnamefont {M.}~\bibnamefont {Wang}}, \bibinfo {author} {\bibfnamefont {S.}~\bibnamefont {Mi}}, \bibinfo {author} {\bibfnamefont {J.}~\bibnamefont {Huang}}, \bibinfo {author} {\bibfnamefont {F.}~\bibnamefont {Pan}}, \bibinfo {author} {\bibfnamefont {X.}~\bibnamefont {Gui}}, \emph {et~al.},\ }\bibfield  {title} {\bibinfo {title} {{F}e{P}d$_2${T}e$_2$: {A}n {A}nisotropic {T}wo-{D}imensional {F}erromagnet with {O}ne-{D}imensional {F}e {C}hains},\ }\href@noop {} {\bibfield  {journal} {\bibinfo  {journal} {Journal of the American Chemical Society}\ }\textbf {\bibinfo {volume} {146}},\ \bibinfo {pages} {21546} (\bibinfo {year} {2024})}\BibitemShut {NoStop}%
\bibitem [{\citenamefont {Yang}\ \emph {et~al.}(1996)\citenamefont {Yang}, \citenamefont {Abell}, \citenamefont {Shang}, \citenamefont {Jones},\ and\ \citenamefont {Gough}}]{yang1996growth}%
  \BibitemOpen
  \bibfield  {author} {\bibinfo {author} {\bibfnamefont {G.}~\bibnamefont {Yang}}, \bibinfo {author} {\bibfnamefont {J.}~\bibnamefont {Abell}}, \bibinfo {author} {\bibfnamefont {P.}~\bibnamefont {Shang}}, \bibinfo {author} {\bibfnamefont {I.}~\bibnamefont {Jones}},\ and\ \bibinfo {author} {\bibfnamefont {C.}~\bibnamefont {Gough}},\ }\bibfield  {title} {\bibinfo {title} {Growth and {M}icrostructure in {B}i$_2${S}r$_2${C}a{C}u$_2${O}$_y$ {S}ingle {C}rystals},\ }\href@noop {} {\bibfield  {journal} {\bibinfo  {journal} {Journal of crystal growth}\ }\textbf {\bibinfo {volume} {166}},\ \bibinfo {pages} {820} (\bibinfo {year} {1996})}\BibitemShut {NoStop}%
\bibitem [{\citenamefont {Murthy}\ and\ \citenamefont {Venimadhav}(2017)}]{murthy20174f}%
  \BibitemOpen
  \bibfield  {author} {\bibinfo {author} {\bibfnamefont {J.~K.}\ \bibnamefont {Murthy}}\ and\ \bibinfo {author} {\bibfnamefont {A.}~\bibnamefont {Venimadhav}},\ }\bibfield  {title} {\bibinfo {title} {4f-3d {E}xchange {C}oupling {I}nduced {E}xchange {B}ias and {F}ield {I}nduced {H}opkinson {P}eak {E}ffects in {G}d$_2${C}o{M}n{O}$_6$},\ }\href@noop {} {\bibfield  {journal} {\bibinfo  {journal} {Journal of Alloys and Compounds}\ }\textbf {\bibinfo {volume} {719}},\ \bibinfo {pages} {341} (\bibinfo {year} {2017})}\BibitemShut {NoStop}%
\bibitem [{\citenamefont {Tang}\ \emph {et~al.}(1994)\citenamefont {Tang}, \citenamefont {Liang}, \citenamefont {Rao},\ and\ \citenamefont {Yan}}]{tang1994study}%
  \BibitemOpen
  \bibfield  {author} {\bibinfo {author} {\bibfnamefont {W.}~\bibnamefont {Tang}}, \bibinfo {author} {\bibfnamefont {J.}~\bibnamefont {Liang}}, \bibinfo {author} {\bibfnamefont {G.}~\bibnamefont {Rao}},\ and\ \bibinfo {author} {\bibfnamefont {X.}~\bibnamefont {Yan}},\ }\bibfield  {title} {\bibinfo {title} {Study of {AC} {S}usceptibility on the {L}a{F}e$_{13-x}${S}i$_x$ {S}ystem},\ }\href@noop {} {\bibfield  {journal} {\bibinfo  {journal} {physica status solidi (a)}\ }\textbf {\bibinfo {volume} {141}},\ \bibinfo {pages} {217} (\bibinfo {year} {1994})}\BibitemShut {NoStop}%
\bibitem [{\citenamefont {Benka}\ \emph {et~al.}(2022)\citenamefont {Benka}, \citenamefont {Bauer}, \citenamefont {Schmakat}, \citenamefont {S{\"a}ubert}, \citenamefont {Seifert}, \citenamefont {Jorba},\ and\ \citenamefont {Pfleiderer}}]{benka2022interplay}%
  \BibitemOpen
  \bibfield  {author} {\bibinfo {author} {\bibfnamefont {G.}~\bibnamefont {Benka}}, \bibinfo {author} {\bibfnamefont {A.}~\bibnamefont {Bauer}}, \bibinfo {author} {\bibfnamefont {P.}~\bibnamefont {Schmakat}}, \bibinfo {author} {\bibfnamefont {S.}~\bibnamefont {S{\"a}ubert}}, \bibinfo {author} {\bibfnamefont {M.}~\bibnamefont {Seifert}}, \bibinfo {author} {\bibfnamefont {P.}~\bibnamefont {Jorba}},\ and\ \bibinfo {author} {\bibfnamefont {C.}~\bibnamefont {Pfleiderer}},\ }\bibfield  {title} {\bibinfo {title} {Interplay of {I}tinerant {M}agnetism and {S}pin-{G}lass {B}ehavior in {F}e$_x${C}r$_{1-x}$},\ }\href@noop {} {\bibfield  {journal} {\bibinfo  {journal} {Physical Review Materials}\ }\textbf {\bibinfo {volume} {6}},\ \bibinfo {pages} {044407} (\bibinfo {year} {2022})}\BibitemShut {NoStop}%
\bibitem [{\citenamefont {Karma}\ \emph {et~al.}(2023)\citenamefont {Karma}, \citenamefont {Harinkhere}, \citenamefont {Karil},\ and\ \citenamefont {Dager}}]{karma2023study}%
  \BibitemOpen
  \bibfield  {author} {\bibinfo {author} {\bibfnamefont {N.}~\bibnamefont {Karma}}, \bibinfo {author} {\bibfnamefont {D.}~\bibnamefont {Harinkhere}}, \bibinfo {author} {\bibfnamefont {P.}~\bibnamefont {Karil}},\ and\ \bibinfo {author} {\bibfnamefont {H.}~\bibnamefont {Dager}},\ }\bibfield  {title} {\bibinfo {title} {Study of {AC} {S}usceptibility of {L}a$_{0.8}${S}r$_{0.2}${M}n{O}$_3$ at {D}ifferent {F}requencies},\ }in\ \href@noop {} {\emph {\bibinfo {booktitle} {Journal of Physics: Conference Series}}},\ Vol.\ \bibinfo {volume} {2603}\ (\bibinfo {organization} {IOP Publishing},\ \bibinfo {year} {2023})\ p.\ \bibinfo {pages} {012030}\BibitemShut {NoStop}%
\bibitem [{\citenamefont {Drobac}(1996)}]{drobac1996critical}%
  \BibitemOpen
  \bibfield  {author} {\bibinfo {author} {\bibfnamefont {{\DJ}.}~\bibnamefont {Drobac}},\ }\bibfield  {title} {\bibinfo {title} {Critical {E}xponents from {H}igh-{P}recision {AC} {S}usceptibility {D}ata},\ }\href@noop {} {\bibfield  {journal} {\bibinfo  {journal} {Journal of magnetism and magnetic materials}\ }\textbf {\bibinfo {volume} {159}},\ \bibinfo {pages} {159} (\bibinfo {year} {1996})}\BibitemShut {NoStop}%
\bibitem [{\citenamefont {Pramanik}\ and\ \citenamefont {Banerjee}(2008)}]{pramanik2008phase}%
  \BibitemOpen
  \bibfield  {author} {\bibinfo {author} {\bibfnamefont {A.}~\bibnamefont {Pramanik}}\ and\ \bibinfo {author} {\bibfnamefont {A.}~\bibnamefont {Banerjee}},\ }\bibfield  {title} {\bibinfo {title} {Phase {S}eparation and the {E}ffect of {Q}uenched {D}isorder in {P}r$_{0.5}${S}r$_{0. 5}${M}n{O}$_3$},\ }\href@noop {} {\bibfield  {journal} {\bibinfo  {journal} {Journal of Physics: Condensed Matter}\ }\textbf {\bibinfo {volume} {20}},\ \bibinfo {pages} {275207} (\bibinfo {year} {2008})}\BibitemShut {NoStop}%
\bibitem [{\citenamefont {Li}\ and\ \citenamefont {Fang}(2012)}]{li2012glassy}%
  \BibitemOpen
  \bibfield  {author} {\bibinfo {author} {\bibfnamefont {F.}~\bibnamefont {Li}}\ and\ \bibinfo {author} {\bibfnamefont {J.}~\bibnamefont {Fang}},\ }\bibfield  {title} {\bibinfo {title} {Glassy-{F}erromagnetic {B}ehavior in {E}u$_{0.5}${S}r$_{0.5}${C}o{O}$_3$},\ }\href@noop {} {\bibfield  {journal} {\bibinfo  {journal} {Journal of magnetism and magnetic materials}\ }\textbf {\bibinfo {volume} {324}},\ \bibinfo {pages} {2664} (\bibinfo {year} {2012})}\BibitemShut {NoStop}%
\bibitem [{\citenamefont {Murthy}\ \emph {et~al.}(2016)\citenamefont {Murthy}, \citenamefont {Chandrasekhar}, \citenamefont {Wu}, \citenamefont {Yang}, \citenamefont {Lin},\ and\ \citenamefont {Venimadhav}}]{murthy2016antisite}%
  \BibitemOpen
  \bibfield  {author} {\bibinfo {author} {\bibfnamefont {J.~K.}\ \bibnamefont {Murthy}}, \bibinfo {author} {\bibfnamefont {K.}~\bibnamefont {Chandrasekhar}}, \bibinfo {author} {\bibfnamefont {H.}~\bibnamefont {Wu}}, \bibinfo {author} {\bibfnamefont {H.}~\bibnamefont {Yang}}, \bibinfo {author} {\bibfnamefont {J.-Y.}\ \bibnamefont {Lin}},\ and\ \bibinfo {author} {\bibfnamefont {A.}~\bibnamefont {Venimadhav}},\ }\bibfield  {title} {\bibinfo {title} {Antisite {D}isorder {D}riven {S}pontaneous {E}xchange {B}ias {E}ffect in {L}a$_{2-x}${S}r$_x${C}o{M}n{O}$_6$ (0$\leq$ x$\leq$ 1)},\ }\href@noop {} {\bibfield  {journal} {\bibinfo  {journal} {Journal of Physics: Condensed Matter}\ }\textbf {\bibinfo {volume} {28}},\ \bibinfo {pages} {086003} (\bibinfo {year} {2016})}\BibitemShut {NoStop}%
\bibitem [{\citenamefont {Sl{\'a}ma}\ \emph {et~al.}(2017)\citenamefont {Sl{\'a}ma}, \citenamefont {U{\v{s}}{\'a}kov{\'a}}, \citenamefont {{\v{S}}oka}, \citenamefont {Dosoudil},\ and\ \citenamefont {Jan{\v{c}}{\'a}rik}}]{slama2017hopkinson}%
  \BibitemOpen
  \bibfield  {author} {\bibinfo {author} {\bibfnamefont {J.}~\bibnamefont {Sl{\'a}ma}}, \bibinfo {author} {\bibfnamefont {M.}~\bibnamefont {U{\v{s}}{\'a}kov{\'a}}}, \bibinfo {author} {\bibfnamefont {M.}~\bibnamefont {{\v{S}}oka}}, \bibinfo {author} {\bibfnamefont {R.}~\bibnamefont {Dosoudil}},\ and\ \bibinfo {author} {\bibfnamefont {V.}~\bibnamefont {Jan{\v{c}}{\'a}rik}},\ }\bibfield  {title} {\bibinfo {title} {Hopkinson {E}ffect in {S}oft and {H}ard {M}agnetic {F}errites},\ }\href@noop {} {\bibfield  {journal} {\bibinfo  {journal} {Acta Physica Polonica A}\ }\textbf {\bibinfo {volume} {131}},\ \bibinfo {pages} {762} (\bibinfo {year} {2017})}\BibitemShut {NoStop}%
\bibitem [{\citenamefont {Bhasker}\ \emph {et~al.}(2018)\citenamefont {Bhasker}, \citenamefont {Choudary},\ and\ \citenamefont {Reddy}}]{bhasker2018modulation}%
  \BibitemOpen
  \bibfield  {author} {\bibinfo {author} {\bibfnamefont {S.~U.}\ \bibnamefont {Bhasker}}, \bibinfo {author} {\bibfnamefont {G.}~\bibnamefont {Choudary}},\ and\ \bibinfo {author} {\bibfnamefont {M.~R.}\ \bibnamefont {Reddy}},\ }\bibfield  {title} {\bibinfo {title} {Modulation in {M}agnetic {E}xchange {I}nteraction, {C}ore {S}hell {S}tructure and {H}opkinson’s {P}eak with {C}hromium {S}ubstitution into {N}i$_{0.75}${C}o$_{0.25}${F}e$_2$o$_4$ {N}ano {P}articles},\ }\href@noop {} {\bibfield  {journal} {\bibinfo  {journal} {Journal of Magnetism and Magnetic Materials}\ }\textbf {\bibinfo {volume} {454}},\ \bibinfo {pages} {349} (\bibinfo {year} {2018})}\BibitemShut {NoStop}%
\bibitem [{\citenamefont {Gaunt}\ \emph {et~al.}(1981)\citenamefont {Gaunt}, \citenamefont {Ho}, \citenamefont {Williams},\ and\ \citenamefont {Cochrane}}]{gaunt1981critical}%
  \BibitemOpen
  \bibfield  {author} {\bibinfo {author} {\bibfnamefont {P.}~\bibnamefont {Gaunt}}, \bibinfo {author} {\bibfnamefont {S.}~\bibnamefont {Ho}}, \bibinfo {author} {\bibfnamefont {G.}~\bibnamefont {Williams}},\ and\ \bibinfo {author} {\bibfnamefont {R.}~\bibnamefont {Cochrane}},\ }\bibfield  {title} {\bibinfo {title} {Critical {B}ehavior of an {A}morphous {F}erromagnet},\ }\href@noop {} {\bibfield  {journal} {\bibinfo  {journal} {Physical Review B}\ }\textbf {\bibinfo {volume} {23}},\ \bibinfo {pages} {251} (\bibinfo {year} {1981})}\BibitemShut {NoStop}%
\bibitem [{\citenamefont {Pippard}(1989)}]{pippard1989magnetoresistance}%
  \BibitemOpen
  \bibfield  {author} {\bibinfo {author} {\bibfnamefont {A.~B.}\ \bibnamefont {Pippard}},\ }\href@noop {} {\emph {\bibinfo {title} {Magnetoresistance in {M}etals}}},\ Vol.~\bibinfo {volume} {2}\ (\bibinfo  {publisher} {Cambridge university press},\ \bibinfo {year} {1989})\BibitemShut {NoStop}%
\bibitem [{\citenamefont {Ramos}\ \emph {et~al.}(2008)\citenamefont {Ramos}, \citenamefont {Arora},\ and\ \citenamefont {Shvets}}]{ramos2008anomalous}%
  \BibitemOpen
  \bibfield  {author} {\bibinfo {author} {\bibfnamefont {R.}~\bibnamefont {Ramos}}, \bibinfo {author} {\bibfnamefont {S.}~\bibnamefont {Arora}},\ and\ \bibinfo {author} {\bibfnamefont {I.}~\bibnamefont {Shvets}},\ }\bibfield  {title} {\bibinfo {title} {Anomalous {A}nisotropic {M}agnetoresistance in {E}pitaxial {F}e$_3${O}$_4$ {T}hin {F}ilms on {M}g{O} (001)},\ }\href@noop {} {\bibfield  {journal} {\bibinfo  {journal} {Physical Review B—Condensed Matter and Materials Physics}\ }\textbf {\bibinfo {volume} {78}},\ \bibinfo {pages} {214402} (\bibinfo {year} {2008})}\BibitemShut {NoStop}%
\bibitem [{\citenamefont {Ando}\ \emph {et~al.}(2003)\citenamefont {Ando}, \citenamefont {Lavrov},\ and\ \citenamefont {Komiya}}]{ando2003anisotropic}%
  \BibitemOpen
  \bibfield  {author} {\bibinfo {author} {\bibfnamefont {Y.}~\bibnamefont {Ando}}, \bibinfo {author} {\bibfnamefont {A.}~\bibnamefont {Lavrov}},\ and\ \bibinfo {author} {\bibfnamefont {S.}~\bibnamefont {Komiya}},\ }\bibfield  {title} {\bibinfo {title} {Anisotropic {M}agnetoresistance in {L}ightly {D}oped {L}a$_{2-x}${S}r$_x${C}u{O}$_4$: {I}mpact of {A}ntiphase {D}omain {B}oundaries on the {E}lectron {T}ransport},\ }\href@noop {} {\bibfield  {journal} {\bibinfo  {journal} {Physical review letters}\ }\textbf {\bibinfo {volume} {90}},\ \bibinfo {pages} {247003} (\bibinfo {year} {2003})}\BibitemShut {NoStop}%
\bibitem [{\citenamefont {Ma}\ \emph {et~al.}(2023)\citenamefont {Ma}, \citenamefont {Huang}, \citenamefont {Wang}, \citenamefont {Liu}, \citenamefont {Zhang}, \citenamefont {Lu},\ and\ \citenamefont {Xiang}}]{ma2023anisotropic}%
  \BibitemOpen
  \bibfield  {author} {\bibinfo {author} {\bibfnamefont {X.}~\bibnamefont {Ma}}, \bibinfo {author} {\bibfnamefont {M.}~\bibnamefont {Huang}}, \bibinfo {author} {\bibfnamefont {S.}~\bibnamefont {Wang}}, \bibinfo {author} {\bibfnamefont {P.}~\bibnamefont {Liu}}, \bibinfo {author} {\bibfnamefont {Y.}~\bibnamefont {Zhang}}, \bibinfo {author} {\bibfnamefont {Y.}~\bibnamefont {Lu}},\ and\ \bibinfo {author} {\bibfnamefont {B.}~\bibnamefont {Xiang}},\ }\bibfield  {title} {\bibinfo {title} {Anisotropic {M}agnetoresistance and {P}lanar {H}all {E}ffect in {L}ayered {R}oom-{T}emperature {F}erromagnet {C}r$_{1.2}${T}e$_{2}$},\ }\href@noop {} {\bibfield  {journal} {\bibinfo  {journal} {ACS Applied Electronic Materials}\ }\textbf {\bibinfo {volume} {5}},\ \bibinfo {pages} {2838} (\bibinfo {year} {2023})}\BibitemShut {NoStop}%
\bibitem [{\citenamefont {Bibes}\ \emph {et~al.}(2000)\citenamefont {Bibes}, \citenamefont {Mart{\i}nez}, \citenamefont {Fontcuberta}, \citenamefont {Trtik}, \citenamefont {Ferrater}, \citenamefont {S{\'a}nchez}, \citenamefont {Varela}, \citenamefont {Hiergeist},\ and\ \citenamefont {Steenbeck}}]{bibes2000anisotropic}%
  \BibitemOpen
  \bibfield  {author} {\bibinfo {author} {\bibfnamefont {M.}~\bibnamefont {Bibes}}, \bibinfo {author} {\bibfnamefont {B.}~\bibnamefont {Mart{\i}nez}}, \bibinfo {author} {\bibfnamefont {J.}~\bibnamefont {Fontcuberta}}, \bibinfo {author} {\bibfnamefont {V.}~\bibnamefont {Trtik}}, \bibinfo {author} {\bibfnamefont {C.}~\bibnamefont {Ferrater}}, \bibinfo {author} {\bibfnamefont {F.}~\bibnamefont {S{\'a}nchez}}, \bibinfo {author} {\bibfnamefont {M.}~\bibnamefont {Varela}}, \bibinfo {author} {\bibfnamefont {R.}~\bibnamefont {Hiergeist}},\ and\ \bibinfo {author} {\bibfnamefont {K.}~\bibnamefont {Steenbeck}},\ }\bibfield  {title} {\bibinfo {title} {Anisotropic {M}agnetoresistance of ($00h$),($0hh$) and ($hhh$) {L}a$_{2/3}${S}r$_{1/3}${M}n{O}$_3$ {T}hin {F}ilms on (001) {S}i {S}ubstrates},\ }\href@noop {} {\bibfield  {journal} {\bibinfo  {journal} {Journal of magnetism and magnetic materials}\ }\textbf {\bibinfo {volume} {211}},\ \bibinfo {pages} {206} (\bibinfo {year} {2000})}\BibitemShut {NoStop}%
\bibitem [{\citenamefont {Bibes}\ \emph {et~al.}(2005)\citenamefont {Bibes}, \citenamefont {Laukhin}, \citenamefont {Valencia}, \citenamefont {Martinez}, \citenamefont {Fontcuberta}, \citenamefont {Gorbenko}, \citenamefont {Kaul},\ and\ \citenamefont {Martinez}}]{bibes2005anisotropic}%
  \BibitemOpen
  \bibfield  {author} {\bibinfo {author} {\bibfnamefont {M.}~\bibnamefont {Bibes}}, \bibinfo {author} {\bibfnamefont {V.}~\bibnamefont {Laukhin}}, \bibinfo {author} {\bibfnamefont {S.}~\bibnamefont {Valencia}}, \bibinfo {author} {\bibfnamefont {B.}~\bibnamefont {Martinez}}, \bibinfo {author} {\bibfnamefont {J.}~\bibnamefont {Fontcuberta}}, \bibinfo {author} {\bibfnamefont {O.~Y.}\ \bibnamefont {Gorbenko}}, \bibinfo {author} {\bibfnamefont {A.}~\bibnamefont {Kaul}},\ and\ \bibinfo {author} {\bibfnamefont {J.}~\bibnamefont {Martinez}},\ }\bibfield  {title} {\bibinfo {title} {Anisotropic {M}agnetoresistance and {A}nomalous {H}all {E}ffect in {M}anganite {T}hin {F}ilms},\ }\href@noop {} {\bibfield  {journal} {\bibinfo  {journal} {Journal of Physics: Condensed Matter}\ }\textbf {\bibinfo {volume} {17}},\ \bibinfo {pages} {2733} (\bibinfo {year} {2005})}\BibitemShut {NoStop}%
\bibitem [{\citenamefont {Dai}\ \emph {et~al.}(2022)\citenamefont {Dai}, \citenamefont {Zhao}, \citenamefont {Ma}, \citenamefont {Tang}, \citenamefont {Qiu}, \citenamefont {Liu}, \citenamefont {Yuan},\ and\ \citenamefont {Zhou}}]{dai2022fourfold}%
  \BibitemOpen
  \bibfield  {author} {\bibinfo {author} {\bibfnamefont {Y.}~\bibnamefont {Dai}}, \bibinfo {author} {\bibfnamefont {Y.}~\bibnamefont {Zhao}}, \bibinfo {author} {\bibfnamefont {L.}~\bibnamefont {Ma}}, \bibinfo {author} {\bibfnamefont {M.}~\bibnamefont {Tang}}, \bibinfo {author} {\bibfnamefont {X.}~\bibnamefont {Qiu}}, \bibinfo {author} {\bibfnamefont {Y.}~\bibnamefont {Liu}}, \bibinfo {author} {\bibfnamefont {Z.}~\bibnamefont {Yuan}},\ and\ \bibinfo {author} {\bibfnamefont {S.}~\bibnamefont {Zhou}},\ }\bibfield  {title} {\bibinfo {title} {Fourfold {A}nisotropic {M}agnetoresistance of {L}1$_{0}${F}e{P}t {D}ue to {R}elaxation {T}ime {A}nisotropy},\ }\href@noop {} {\bibfield  {journal} {\bibinfo  {journal} {Physical Review Letters}\ }\textbf {\bibinfo {volume} {128}},\ \bibinfo {pages} {247202} (\bibinfo {year} {2022})}\BibitemShut {NoStop}%
\bibitem [{\citenamefont {Eerenstein}\ \emph {et~al.}(2002)\citenamefont {Eerenstein}, \citenamefont {Palstra}, \citenamefont {Saxena},\ and\ \citenamefont {Hibma}}]{eerenstein2002spin}%
  \BibitemOpen
  \bibfield  {author} {\bibinfo {author} {\bibfnamefont {W.}~\bibnamefont {Eerenstein}}, \bibinfo {author} {\bibfnamefont {T.}~\bibnamefont {Palstra}}, \bibinfo {author} {\bibfnamefont {S.}~\bibnamefont {Saxena}},\ and\ \bibinfo {author} {\bibfnamefont {T.}~\bibnamefont {Hibma}},\ }\bibfield  {title} {\bibinfo {title} {Spin-{P}olarized {T}ransport across {S}harp {A}ntiferromagnetic {B}oundaries},\ }\href@noop {} {\bibfield  {journal} {\bibinfo  {journal} {Physical review letters}\ }\textbf {\bibinfo {volume} {88}},\ \bibinfo {pages} {247204} (\bibinfo {year} {2002})}\BibitemShut {NoStop}%
\bibitem [{\citenamefont {Li}\ \emph {et~al.}(2010)\citenamefont {Li}, \citenamefont {Jin}, \citenamefont {Jiang},\ and\ \citenamefont {Bai}}]{li2010origin}%
  \BibitemOpen
  \bibfield  {author} {\bibinfo {author} {\bibfnamefont {P.}~\bibnamefont {Li}}, \bibinfo {author} {\bibfnamefont {C.}~\bibnamefont {Jin}}, \bibinfo {author} {\bibfnamefont {E.}~\bibnamefont {Jiang}},\ and\ \bibinfo {author} {\bibfnamefont {H.}~\bibnamefont {Bai}},\ }\bibfield  {title} {\bibinfo {title} {Origin of the {T}wofold and {F}ourfold {S}ymmetric {A}nisotropic {M}agnetoresistance in {E}pitaxial {F}e$_{3}${O}$_{4}$ {F}ilms},\ }\href@noop {} {\bibfield  {journal} {\bibinfo  {journal} {Journal of Applied Physics}\ }\textbf {\bibinfo {volume} {108}} (\bibinfo {year} {2010})}\BibitemShut {NoStop}%
\bibitem [{\citenamefont {Margulies}\ \emph {et~al.}(1997)\citenamefont {Margulies}, \citenamefont {Parker}, \citenamefont {Rudee}, \citenamefont {Spada}, \citenamefont {Chapman}, \citenamefont {Aitchison},\ and\ \citenamefont {Berkowitz}}]{margulies1997origin}%
  \BibitemOpen
  \bibfield  {author} {\bibinfo {author} {\bibfnamefont {D.}~\bibnamefont {Margulies}}, \bibinfo {author} {\bibfnamefont {F.}~\bibnamefont {Parker}}, \bibinfo {author} {\bibfnamefont {M.}~\bibnamefont {Rudee}}, \bibinfo {author} {\bibfnamefont {F.}~\bibnamefont {Spada}}, \bibinfo {author} {\bibfnamefont {J.}~\bibnamefont {Chapman}}, \bibinfo {author} {\bibfnamefont {P.}~\bibnamefont {Aitchison}},\ and\ \bibinfo {author} {\bibfnamefont {A.}~\bibnamefont {Berkowitz}},\ }\bibfield  {title} {\bibinfo {title} {Origin of the {A}nomalous {M}agnetic {B}ehavior in {S}ingle {C}rystal {F}e$_3${O}$_4$ {F}ilms},\ }\href@noop {} {\bibfield  {journal} {\bibinfo  {journal} {Physical Review Letters}\ }\textbf {\bibinfo {volume} {79}},\ \bibinfo {pages} {5162} (\bibinfo {year} {1997})}\BibitemShut {NoStop}%
\bibitem [{\citenamefont {Arrott}\ and\ \citenamefont {Noakes}(1967)}]{arrott1967approximate}%
  \BibitemOpen
  \bibfield  {author} {\bibinfo {author} {\bibfnamefont {A.}~\bibnamefont {Arrott}}\ and\ \bibinfo {author} {\bibfnamefont {J.~E.}\ \bibnamefont {Noakes}},\ }\bibfield  {title} {\bibinfo {title} {Approximate {E}quation of {S}tate for {N}ickel near {I}ts {C}ritical {T}emperature},\ }\href@noop {} {\bibfield  {journal} {\bibinfo  {journal} {Physical Review Letters}\ }\textbf {\bibinfo {volume} {19}},\ \bibinfo {pages} {786} (\bibinfo {year} {1967})}\BibitemShut {NoStop}%
\bibitem [{\citenamefont {Fisher}(1967)}]{fisher1967theory}%
  \BibitemOpen
  \bibfield  {author} {\bibinfo {author} {\bibfnamefont {M.~E.}\ \bibnamefont {Fisher}},\ }\bibfield  {title} {\bibinfo {title} {The {T}heory of {E}quilibrium {C}ritical {P}henomena},\ }\href@noop {} {\bibfield  {journal} {\bibinfo  {journal} {Reports on progress in physics}\ }\textbf {\bibinfo {volume} {30}},\ \bibinfo {pages} {615} (\bibinfo {year} {1967})}\BibitemShut {NoStop}%
\bibitem [{\citenamefont {Wang}\ \emph {et~al.}(2022)\citenamefont {Wang}, \citenamefont {Bedoya-Pinto}, \citenamefont {Blei}, \citenamefont {Dismukes}, \citenamefont {Hamo}, \citenamefont {Jenkins}, \citenamefont {Koperski}, \citenamefont {Liu}, \citenamefont {Sun}, \citenamefont {Telford} \emph {et~al.}}]{wang2022magnetic}%
  \BibitemOpen
  \bibfield  {author} {\bibinfo {author} {\bibfnamefont {Q.~H.}\ \bibnamefont {Wang}}, \bibinfo {author} {\bibfnamefont {A.}~\bibnamefont {Bedoya-Pinto}}, \bibinfo {author} {\bibfnamefont {M.}~\bibnamefont {Blei}}, \bibinfo {author} {\bibfnamefont {A.~H.}\ \bibnamefont {Dismukes}}, \bibinfo {author} {\bibfnamefont {A.}~\bibnamefont {Hamo}}, \bibinfo {author} {\bibfnamefont {S.}~\bibnamefont {Jenkins}}, \bibinfo {author} {\bibfnamefont {M.}~\bibnamefont {Koperski}}, \bibinfo {author} {\bibfnamefont {Y.}~\bibnamefont {Liu}}, \bibinfo {author} {\bibfnamefont {Q.-C.}\ \bibnamefont {Sun}}, \bibinfo {author} {\bibfnamefont {E.~J.}\ \bibnamefont {Telford}}, \emph {et~al.},\ }\bibfield  {title} {\bibinfo {title} {The {M}agnetic {G}enome of {T}wo-{D}imensional van der {W}aals {M}aterials},\ }\href@noop {} {\bibfield  {journal} {\bibinfo  {journal} {ACS nano}\ }\textbf {\bibinfo {volume} {16}},\ \bibinfo {pages} {6960} (\bibinfo {year} {2022})}\BibitemShut {NoStop}%
\bibitem [{\citenamefont {Pramanik}\ and\ \citenamefont {Banerjee}(2009)}]{pramanik2009critical}%
  \BibitemOpen
  \bibfield  {author} {\bibinfo {author} {\bibfnamefont {A.}~\bibnamefont {Pramanik}}\ and\ \bibinfo {author} {\bibfnamefont {A.}~\bibnamefont {Banerjee}},\ }\bibfield  {title} {\bibinfo {title} {Critical {B}ehavior at {P}aramagnetic to {F}erromagnetic {P}hase {T}ransition in {P}r$_{0.5}$ {S}r$_{0.5}${M}n{O}$_3$: {A} {B}ulk {M}agnetization {S}tudy},\ }\href@noop {} {\bibfield  {journal} {\bibinfo  {journal} {Physical Review B—Condensed Matter and Materials Physics}\ }\textbf {\bibinfo {volume} {79}},\ \bibinfo {pages} {214426} (\bibinfo {year} {2009})}\BibitemShut {NoStop}%
\bibitem [{\citenamefont {Kouvel}\ and\ \citenamefont {Fisher}(1964)}]{kouvel1964detailed}%
  \BibitemOpen
  \bibfield  {author} {\bibinfo {author} {\bibfnamefont {J.~S.}\ \bibnamefont {Kouvel}}\ and\ \bibinfo {author} {\bibfnamefont {M.~E.}\ \bibnamefont {Fisher}},\ }\bibfield  {title} {\bibinfo {title} {Detailed {M}agnetic {B}ehavior of {N}ickel near {I}ts {C}urie {P}oint},\ }\href@noop {} {\bibfield  {journal} {\bibinfo  {journal} {Physical Review}\ }\textbf {\bibinfo {volume} {136}},\ \bibinfo {pages} {A1626} (\bibinfo {year} {1964})}\BibitemShut {NoStop}%
\bibitem [{\citenamefont {Fisher}\ \emph {et~al.}(1972)\citenamefont {Fisher}, \citenamefont {Ma},\ and\ \citenamefont {Nickel}}]{fisher1972critical}%
  \BibitemOpen
  \bibfield  {author} {\bibinfo {author} {\bibfnamefont {M.~E.}\ \bibnamefont {Fisher}}, \bibinfo {author} {\bibfnamefont {S.-k.}\ \bibnamefont {Ma}},\ and\ \bibinfo {author} {\bibfnamefont {B.}~\bibnamefont {Nickel}},\ }\bibfield  {title} {\bibinfo {title} {Critical {E}xponents for {L}ong-{R}ange {I}nteractions},\ }\href@noop {} {\bibfield  {journal} {\bibinfo  {journal} {Physical Review Letters}\ }\textbf {\bibinfo {volume} {29}},\ \bibinfo {pages} {917} (\bibinfo {year} {1972})}\BibitemShut {NoStop}%
\bibitem [{\citenamefont {Omrani}\ \emph {et~al.}(2016)\citenamefont {Omrani}, \citenamefont {Mansouri}, \citenamefont {Koubaa}, \citenamefont {Koubaa},\ and\ \citenamefont {Cheikhrouhou}}]{omrani2016critical}%
  \BibitemOpen
  \bibfield  {author} {\bibinfo {author} {\bibfnamefont {H.}~\bibnamefont {Omrani}}, \bibinfo {author} {\bibfnamefont {M.}~\bibnamefont {Mansouri}}, \bibinfo {author} {\bibfnamefont {W.~C.}\ \bibnamefont {Koubaa}}, \bibinfo {author} {\bibfnamefont {M.}~\bibnamefont {Koubaa}},\ and\ \bibinfo {author} {\bibfnamefont {A.}~\bibnamefont {Cheikhrouhou}},\ }\bibfield  {title} {\bibinfo {title} {Critical {B}ehavior {S}tudy near the {P}aramagnetic to {F}erromagnetic {P}hase {T}ransition {T}emperature in {P}r$_{0.6-x}${E}r$_x${C}a$_{0.1}${S}r$_{0.3}${M}n{O}$_3$ (x= 0, 0.02 and 0.06) {M}anganites},\ }\href@noop {} {\bibfield  {journal} {\bibinfo  {journal} {RSC Advances}\ }\textbf {\bibinfo {volume} {6}},\ \bibinfo {pages} {78017} (\bibinfo {year} {2016})}\BibitemShut {NoStop}%
\bibitem [{\citenamefont {Chauhan}\ \emph {et~al.}(2022)\citenamefont {Chauhan}, \citenamefont {Kumar}, \citenamefont {Tiwari}, \citenamefont {Tiwari},\ and\ \citenamefont {Ghosh}}]{chauhan2022different}%
  \BibitemOpen
  \bibfield  {author} {\bibinfo {author} {\bibfnamefont {H.~C.}\ \bibnamefont {Chauhan}}, \bibinfo {author} {\bibfnamefont {B.}~\bibnamefont {Kumar}}, \bibinfo {author} {\bibfnamefont {A.}~\bibnamefont {Tiwari}}, \bibinfo {author} {\bibfnamefont {J.~K.}\ \bibnamefont {Tiwari}},\ and\ \bibinfo {author} {\bibfnamefont {S.}~\bibnamefont {Ghosh}},\ }\bibfield  {title} {\bibinfo {title} {Different {C}ritical {E}xponenxts on {T}wo {S}ides of a {T}ransition: {O}bservation of {C}rossover from {I}sing to {H}eisenberg {E}xchange in {S}kyrmion {H}ost {C}u$_2${OS}e{O}$_3$},\ }\href@noop {} {\bibfield  {journal} {\bibinfo  {journal} {Physical Review Letters}\ }\textbf {\bibinfo {volume} {128}},\ \bibinfo {pages} {015703} (\bibinfo {year} {2022})}\BibitemShut {NoStop}%
\bibitem [{\citenamefont {Zhang}\ \emph {et~al.}(2016)\citenamefont {Zhang}, \citenamefont {Han}, \citenamefont {Ge}, \citenamefont {Du}, \citenamefont {Jin}, \citenamefont {Wei}, \citenamefont {Fan}, \citenamefont {Zhang}, \citenamefont {Pi},\ and\ \citenamefont {Zhang}}]{zhang2016critical}%
  \BibitemOpen
  \bibfield  {author} {\bibinfo {author} {\bibfnamefont {L.}~\bibnamefont {Zhang}}, \bibinfo {author} {\bibfnamefont {H.}~\bibnamefont {Han}}, \bibinfo {author} {\bibfnamefont {M.}~\bibnamefont {Ge}}, \bibinfo {author} {\bibfnamefont {H.}~\bibnamefont {Du}}, \bibinfo {author} {\bibfnamefont {C.}~\bibnamefont {Jin}}, \bibinfo {author} {\bibfnamefont {W.}~\bibnamefont {Wei}}, \bibinfo {author} {\bibfnamefont {J.}~\bibnamefont {Fan}}, \bibinfo {author} {\bibfnamefont {C.}~\bibnamefont {Zhang}}, \bibinfo {author} {\bibfnamefont {L.}~\bibnamefont {Pi}},\ and\ \bibinfo {author} {\bibfnamefont {Y.}~\bibnamefont {Zhang}},\ }\bibfield  {title} {\bibinfo {title} {Critical {P}henomenon of the {N}ear {R}oom {T}emperature {S}kyrmion {M}aterial {F}e{G}e},\ }\href@noop {} {\bibfield  {journal} {\bibinfo  {journal} {Scientific reports}\ }\textbf {\bibinfo {volume} {6}},\ \bibinfo {pages} {22397} (\bibinfo {year} {2016})}\BibitemShut {NoStop}%
\bibitem [{\citenamefont {Nag}\ \emph {et~al.}(2019)\citenamefont {Nag}, \citenamefont {Sarkar}, \citenamefont {Pal},\ and\ \citenamefont {Bose}}]{nag2019magnetic}%
  \BibitemOpen
  \bibfield  {author} {\bibinfo {author} {\bibfnamefont {R.}~\bibnamefont {Nag}}, \bibinfo {author} {\bibfnamefont {B.}~\bibnamefont {Sarkar}}, \bibinfo {author} {\bibfnamefont {S.}~\bibnamefont {Pal}},\ and\ \bibinfo {author} {\bibfnamefont {E.}~\bibnamefont {Bose}},\ }\bibfield  {title} {\bibinfo {title} {Magnetic {P}roperties and {C}ritical {B}ehavior of {E}lectron {D}oped {P}olycrystalline {C}a$_{0.85}${S}m$_{0.15}${M}n{O}$_3$ {M}anganite},\ }\href@noop {} {\bibfield  {journal} {\bibinfo  {journal} {Physica Scripta}\ }\textbf {\bibinfo {volume} {94}},\ \bibinfo {pages} {095801} (\bibinfo {year} {2019})}\BibitemShut {NoStop}%
\bibitem [{\citenamefont {Sarkar}\ \emph {et~al.}(2021)\citenamefont {Sarkar}, \citenamefont {Nag},\ and\ \citenamefont {Pal}}]{sarkar2021manifestation}%
  \BibitemOpen
  \bibfield  {author} {\bibinfo {author} {\bibfnamefont {B.}~\bibnamefont {Sarkar}}, \bibinfo {author} {\bibfnamefont {R.}~\bibnamefont {Nag}},\ and\ \bibinfo {author} {\bibfnamefont {S.}~\bibnamefont {Pal}},\ }\bibfield  {title} {\bibinfo {title} {Manifestation of {N}onconventional {M}agnetic {I}nteraction in {E}lectron-{D}oped {C}a$_{0.85}${D}y$_{0.15}${M}n{O}$_3$ from {M}agnetic {P}roperties and {C}ritical {B}ehavior {A}nalysis},\ }\href@noop {} {\bibfield  {journal} {\bibinfo  {journal} {Journal of Superconductivity and Novel Magnetism}\ }\textbf {\bibinfo {volume} {34}},\ \bibinfo {pages} {2059} (\bibinfo {year} {2021})}\BibitemShut {NoStop}%
\bibitem [{\citenamefont {Kundu}\ and\ \citenamefont {Nath}(2013)}]{kundu2013critical}%
  \BibitemOpen
  \bibfield  {author} {\bibinfo {author} {\bibfnamefont {S.}~\bibnamefont {Kundu}}\ and\ \bibinfo {author} {\bibfnamefont {T.}~\bibnamefont {Nath}},\ }\bibfield  {title} {\bibinfo {title} {Critical {B}ehavior and {M}agnetic {R}elaxation {D}ynamics of {N}d$_{0.4}${S}r$_{0.6}${M}n{O}$_3$ {N}anoparticles},\ }\href@noop {} {\bibfield  {journal} {\bibinfo  {journal} {Philosophical Magazine}\ }\textbf {\bibinfo {volume} {93}},\ \bibinfo {pages} {2527} (\bibinfo {year} {2013})}\BibitemShut {NoStop}%
\bibitem [{\citenamefont {Liu}\ \emph {et~al.}(2019)\citenamefont {Liu}, \citenamefont {Wang}, \citenamefont {Fan}, \citenamefont {Pi}, \citenamefont {Ge}, \citenamefont {Zhang},\ and\ \citenamefont {Zhang}}]{liu2019field}%
  \BibitemOpen
  \bibfield  {author} {\bibinfo {author} {\bibfnamefont {W.}~\bibnamefont {Liu}}, \bibinfo {author} {\bibfnamefont {Y.}~\bibnamefont {Wang}}, \bibinfo {author} {\bibfnamefont {J.}~\bibnamefont {Fan}}, \bibinfo {author} {\bibfnamefont {L.}~\bibnamefont {Pi}}, \bibinfo {author} {\bibfnamefont {M.}~\bibnamefont {Ge}}, \bibinfo {author} {\bibfnamefont {L.}~\bibnamefont {Zhang}},\ and\ \bibinfo {author} {\bibfnamefont {Y.}~\bibnamefont {Zhang}},\ }\bibfield  {title} {\bibinfo {title} {Field-{D}ependent {A}nisotropic {M}agnetic {C}oupling in {L}ayered {F}erromagnetic {F}e$_{3-x}$ {G}e{T}e$_2$},\ }\href@noop {} {\bibfield  {journal} {\bibinfo  {journal} {Physical Review B}\ }\textbf {\bibinfo {volume} {100}},\ \bibinfo {pages} {104403} (\bibinfo {year} {2019})}\BibitemShut {NoStop}%
\bibitem [{\citenamefont {Rumpf}\ \emph {et~al.}(2010)\citenamefont {Rumpf}, \citenamefont {Granitzer},\ and\ \citenamefont {P{\"o}lt}}]{rumpf2010non}%
  \BibitemOpen
  \bibfield  {author} {\bibinfo {author} {\bibfnamefont {K.}~\bibnamefont {Rumpf}}, \bibinfo {author} {\bibfnamefont {P.}~\bibnamefont {Granitzer}},\ and\ \bibinfo {author} {\bibfnamefont {P.}~\bibnamefont {P{\"o}lt}},\ }\bibfield  {title} {\bibinfo {title} {Non-{S}aturating {M}agnetic {B}ehaviour of a {F}erromagnetic {S}emiconductor/{M}etal {N}anocomposite},\ }\href@noop {} {\bibfield  {journal} {\bibinfo  {journal} {Journal of magnetism and magnetic materials}\ }\textbf {\bibinfo {volume} {322}},\ \bibinfo {pages} {1283} (\bibinfo {year} {2010})}\BibitemShut {NoStop}%
\bibitem [{\citenamefont {Kumar}\ and\ \citenamefont {Banerjee}(2012)}]{kumar2012critical}%
  \BibitemOpen
  \bibfield  {author} {\bibinfo {author} {\bibfnamefont {D.}~\bibnamefont {Kumar}}\ and\ \bibinfo {author} {\bibfnamefont {A.}~\bibnamefont {Banerjee}},\ }\bibfield  {title} {\bibinfo {title} {A {C}ritical {E}xamination of {M}agnetic {S}tates of {L}a$_{0.5}${B}a$_{0.5}${C}o{O}$_3$: {N}on-{G}riffiths {P}hase and {I}nteracting {F}erromagnetic-{C}lusters},\ }\href@noop {} {\bibfield  {journal} {\bibinfo  {journal} {arXiv preprint arXiv:1211.4936}\ } (\bibinfo {year} {2012})}\BibitemShut {NoStop}%
\bibitem [{\citenamefont {Zhang}\ \emph {et~al.}(2022)\citenamefont {Zhang}, \citenamefont {Guo}, \citenamefont {Wu}, \citenamefont {Wen}, \citenamefont {Yang}, \citenamefont {Jin}, \citenamefont {Zhang},\ and\ \citenamefont {Chang}}]{zhang2022above}%
  \BibitemOpen
  \bibfield  {author} {\bibinfo {author} {\bibfnamefont {G.}~\bibnamefont {Zhang}}, \bibinfo {author} {\bibfnamefont {F.}~\bibnamefont {Guo}}, \bibinfo {author} {\bibfnamefont {H.}~\bibnamefont {Wu}}, \bibinfo {author} {\bibfnamefont {X.}~\bibnamefont {Wen}}, \bibinfo {author} {\bibfnamefont {L.}~\bibnamefont {Yang}}, \bibinfo {author} {\bibfnamefont {W.}~\bibnamefont {Jin}}, \bibinfo {author} {\bibfnamefont {W.}~\bibnamefont {Zhang}},\ and\ \bibinfo {author} {\bibfnamefont {H.}~\bibnamefont {Chang}},\ }\bibfield  {title} {\bibinfo {title} {Above-{R}oom-{T}emperature {S}trong {I}ntrinsic {F}erromagnetism in 2d van der {W}aals {F}e$_3${G}a{T}e$_2$ with {L}arge {P}erpendicular {M}agnetic {A}nisotropy},\ }\href@noop {} {\bibfield  {journal} {\bibinfo  {journal} {Nature Communications}\ }\textbf {\bibinfo {volume} {13}},\ \bibinfo {pages} {5067} (\bibinfo {year} {2022})}\BibitemShut {NoStop}%
\bibitem [{\citenamefont {Mermin}\ and\ \citenamefont {Wagner}(1966)}]{mermin1966absence}%
  \BibitemOpen
  \bibfield  {author} {\bibinfo {author} {\bibfnamefont {N.~D.}\ \bibnamefont {Mermin}}\ and\ \bibinfo {author} {\bibfnamefont {H.}~\bibnamefont {Wagner}},\ }\bibfield  {title} {\bibinfo {title} {Absence of {F}erromagnetism or {A}ntiferromagnetism in {O}ne-or {T}wo-{D}imensional {I}sotropic {H}eisenberg {M}odels},\ }\href@noop {} {\bibfield  {journal} {\bibinfo  {journal} {Physical Review Letters}\ }\textbf {\bibinfo {volume} {17}},\ \bibinfo {pages} {1133} (\bibinfo {year} {1966})}\BibitemShut {NoStop}%
\bibitem [{\citenamefont {Duan}\ \emph {et~al.}(2022)\citenamefont {Duan}, \citenamefont {Wang}, \citenamefont {Zhang}, \citenamefont {Hu}, \citenamefont {Zhao}, \citenamefont {Feng}, \citenamefont {Zhang}, \citenamefont {Lin}, \citenamefont {Chen}, \citenamefont {Wu} \emph {et~al.}}]{duan2022synthesis}%
  \BibitemOpen
  \bibfield  {author} {\bibinfo {author} {\bibfnamefont {L.}~\bibnamefont {Duan}}, \bibinfo {author} {\bibfnamefont {X.}~\bibnamefont {Wang}}, \bibinfo {author} {\bibfnamefont {J.}~\bibnamefont {Zhang}}, \bibinfo {author} {\bibfnamefont {Z.}~\bibnamefont {Hu}}, \bibinfo {author} {\bibfnamefont {J.}~\bibnamefont {Zhao}}, \bibinfo {author} {\bibfnamefont {Y.}~\bibnamefont {Feng}}, \bibinfo {author} {\bibfnamefont {H.}~\bibnamefont {Zhang}}, \bibinfo {author} {\bibfnamefont {H.-J.}\ \bibnamefont {Lin}}, \bibinfo {author} {\bibfnamefont {C.}~\bibnamefont {Chen}}, \bibinfo {author} {\bibfnamefont {W.}~\bibnamefont {Wu}}, \emph {et~al.},\ }\bibfield  {title} {\bibinfo {title} {Synthesis, {S}tructure, and {M}agnetism in the {F}erromagnet {L}a$_3${M}n{A}s$_5$: {W}ell-{S}eparated {S}pin {C}hains {C}oupled via {I}tinerant {E}lectrons},\ }\href@noop {} {\bibfield  {journal} {\bibinfo  {journal} {Physical Review B}\ }\textbf {\bibinfo {volume} {106}},\ \bibinfo {pages} {184405} (\bibinfo {year} {2022})}\BibitemShut
  {NoStop}%
\bibitem [{\citenamefont {Zhang}\ and\ \citenamefont {Willis}(2001)}]{zhang2001thickness}%
  \BibitemOpen
  \bibfield  {author} {\bibinfo {author} {\bibfnamefont {R.}~\bibnamefont {Zhang}}\ and\ \bibinfo {author} {\bibfnamefont {R.~F.}\ \bibnamefont {Willis}},\ }\bibfield  {title} {\bibinfo {title} {Thickness-{D}ependent {C}urie {T}emperatures of {U}ltrathin {M}agnetic {F}ilms: {E}ffect of the {R}ange of {S}pin-spin {I}nteractions},\ }\href@noop {} {\bibfield  {journal} {\bibinfo  {journal} {Physical review letters}\ }\textbf {\bibinfo {volume} {86}},\ \bibinfo {pages} {2665} (\bibinfo {year} {2001})}\BibitemShut {NoStop}%
\bibitem [{\citenamefont {Deng}\ \emph {et~al.}(2018)\citenamefont {Deng}, \citenamefont {Yu}, \citenamefont {Song}, \citenamefont {Zhang}, \citenamefont {Wang}, \citenamefont {Sun}, \citenamefont {Yi}, \citenamefont {Wu}, \citenamefont {Wu}, \citenamefont {Zhu} \emph {et~al.}}]{deng2018gate}%
  \BibitemOpen
  \bibfield  {author} {\bibinfo {author} {\bibfnamefont {Y.}~\bibnamefont {Deng}}, \bibinfo {author} {\bibfnamefont {Y.}~\bibnamefont {Yu}}, \bibinfo {author} {\bibfnamefont {Y.}~\bibnamefont {Song}}, \bibinfo {author} {\bibfnamefont {J.}~\bibnamefont {Zhang}}, \bibinfo {author} {\bibfnamefont {N.~Z.}\ \bibnamefont {Wang}}, \bibinfo {author} {\bibfnamefont {Z.}~\bibnamefont {Sun}}, \bibinfo {author} {\bibfnamefont {Y.}~\bibnamefont {Yi}}, \bibinfo {author} {\bibfnamefont {Y.~Z.}\ \bibnamefont {Wu}}, \bibinfo {author} {\bibfnamefont {S.}~\bibnamefont {Wu}}, \bibinfo {author} {\bibfnamefont {J.}~\bibnamefont {Zhu}}, \emph {et~al.},\ }\bibfield  {title} {\bibinfo {title} {Gate-{T}unable {R}oom-{T}emperature {F}erromagnetism in {T}wo-{D}imensional {F}e$_3${G}e{T}e$_2$},\ }\href@noop {} {\bibfield  {journal} {\bibinfo  {journal} {Nature}\ }\textbf {\bibinfo {volume} {563}},\ \bibinfo {pages} {94} (\bibinfo {year} {2018})}\BibitemShut {NoStop}%
\bibitem [{\citenamefont {Chen}\ \emph {et~al.}(2024)\citenamefont {Chen}, \citenamefont {Yang}, \citenamefont {Ying},\ and\ \citenamefont {Guo}}]{chen2024high}%
  \BibitemOpen
  \bibfield  {author} {\bibinfo {author} {\bibfnamefont {Z.}~\bibnamefont {Chen}}, \bibinfo {author} {\bibfnamefont {Y.}~\bibnamefont {Yang}}, \bibinfo {author} {\bibfnamefont {T.}~\bibnamefont {Ying}},\ and\ \bibinfo {author} {\bibfnamefont {J.-G.}\ \bibnamefont {Guo}},\ }\bibfield  {title} {\bibinfo {title} {High-${T}_c$ {F}erromagnetic {S}emiconductor in {T}hinned 3d {I}sing {F}erromagnetic {M}etal {F}e$_3${G}a{T}e$_2$},\ }\href@noop {} {\bibfield  {journal} {\bibinfo  {journal} {Nano Letters}\ }\textbf {\bibinfo {volume} {24}},\ \bibinfo {pages} {993} (\bibinfo {year} {2024})}\BibitemShut {NoStop}%
\end{thebibliography}%
\end{document}